\documentclass[prb,twocolumn,showpacs,preprintnumbers,amsmath,amssymb,superscriptaddress,floatfix]{revtex4-1}
\usepackage{graphicx}
\usepackage{dcolumn}
\usepackage{bm}
\usepackage{color}
\usepackage{hyperref}

\begin{document}

\title{Controlling circular polarization of light emitted by quantum dots using chiral photonic crystal slab}

\author{S. V. Lobanov}
\affiliation{A. M. Prokhorov General Physics Institute, Russian Academy of Sciences, Vavilova Street 38, Moscow 119991, Russia}
\affiliation{School of Physics and Astronomy, Cardiff University, Cardiff CF24 3AA, United Kingdom}
\affiliation{Skolkovo Institute of Science and Technology, Novaya Street 100, Skolkovo 143025, Russia}

\author{S. G. Tikhodeev}
\affiliation{A. M. Prokhorov General Physics Institute, Russian Academy of Sciences, Vavilova Street 38, Moscow 119991, Russia}
\affiliation{Institute of Solid State Physics, Russian Academy of Science, Chernogolovka 142432, Russia}
\affiliation{M. V. Lomonosov Moscow State University, Leninskie Gory 1, Moscow 119991, Russia}

\author{N. A. Gippius}
\affiliation{Skolkovo Institute of Science and Technology, Novaya Street 100, Skolkovo 143025, Russia}
\affiliation{A. M. Prokhorov General Physics Institute, Russian Academy of Sciences, Vavilova Street 38, Moscow 119991, Russia}

\author{A. A. Maksimov}
\affiliation{Institute of Solid State Physics, Russian Academy of Science, Chernogolovka 142432, Russia}
\author{E.\ V.\ Filatov}
\affiliation{Institute of Solid State Physics, Russian Academy of Science, Chernogolovka 142432, Russia}
\author{I.~I.~Tartakovskii}
\affiliation{Institute of Solid State Physics, Russian Academy of Science, Chernogolovka 142432, Russia}
\author{V. D. Kulakovskii}
\affiliation{Institute of Solid State Physics, Russian Academy of Science, Chernogolovka 142432, Russia}

\author{T. Weiss}
\affiliation{4$^\mathrm{th}$ Physics Institute and Research Center SCoPE, University of Stuttgart, Stuttgart D-70550, Germany}

\author{C.~Schneider}
\affiliation{Technische Physik, Physikalisches Institut and Wilhelm Conrad R\"ontgen Research Center for Complex Material Systems,
	Universit\"at W\"urzburg, D-97074 W\"urzburg, Germany}
\author{J. Ge{\ss}ler}
\affiliation{Technische Physik, Physikalisches Institut and Wilhelm Conrad R\"ontgen Research Center for Complex Material Systems,
	Universit\"at W\"urzburg, D-97074 W\"urzburg, Germany}
\author{M.\ Kamp}
\affiliation{Technische Physik, Physikalisches Institut and Wilhelm Conrad R\"ontgen Research Center for Complex Material Systems,
	Universit\"at W\"urzburg, D-97074 W\"urzburg, Germany}
\author{S. H\"ofling}
\affiliation{Technische Physik, Physikalisches Institut and Wilhelm Conrad R\"ontgen Research Center for Complex Material Systems,
	Universit\"at W\"urzburg, D-97074 W\"urzburg, Germany}
\affiliation{SUPA, School of Physics and Astronomy, University of St Andrews, St Andrews, KY16 9SS, United Kingdom}



\date{25 September, 2015}
\pacs{71.36.+c, 42.65.Pc, 42.55.Sa}

\begin{abstract}
We study the polarization properties of light emitted by quantum dots that are embedded in chiral photonic crystal structures
made of achiral planar GaAs waveguides. A modification of the electromagnetic mode structure
due to the chiral grating fabricated by partial etching of the wave\-guide layer has been shown to result in a high
 circular polarization degree $\rho_c$ of the quantum dot emission in the absence of external magnetic field. The physical
 nature of the phenomenon can be understood in terms of the reciprocity principle taking into account the structural symmetry. At the
 resonance wavelength, the magnitude of  $|\rho_c|$ is predicted to exceed 98\%. The experimentally achieved
 value of  $|\rho_c|=81$\% is  smaller,  which is due to the contribution of unpolarized light
 scattered by grating defects, thus breaking its periodicity. The achieved polarization degree estimated removing
the unpolarized nonresonant background from the emission spectra
 can be estimated to be as high as 96\%, close to  the theoretical prediction.
\end{abstract}

\maketitle

\section{\label{Sec1} Introduction}

The possibility to control the polarization state of radiation from quantum emitters has drawn attention of researchers
in recent years as it opens various important applications in spin-optoelectronics, quantum information technology, chiral
synthesis and sensing.  A common way for polarization conversion and rotation is the use of wave plates of birefringent
materials and optical gratings. An imbalance between the left- and right-circularly polarized photons
takes place in chiral materials with nonequivalent left- and right-circularly polarized electromagnetic field
modes.\cite{ Stegemeyer1979, Schmidtke2003, Woon2005, DeLeon2015} The imbalance in semiconductors is usually reached by
applying a static magnetic field that disturbs the time-reversal symmetry and leads to a splitting of the left- and right-circularly polarized modes.

Advances in nanoscale fabrication have made it possible to realize artificial photonic structures with desired symmetry and
 density of the environmentally allowed electromagnetic modes, which
allows to control the spontaneous emission rate of
 emitted light, its radiation pattern and direction.
Moreover, the fabrication of chiral nanostructures from achiral semiconductor materials opens the
 possibility to control the polarization of emitted light,\cite{Konishi2011,Shitrit2013}
as the polarization effects are determined
 by the overall structure symmetry.\cite{ Barron1972} This method has considerable advantages in comparison with the others
 as it is compatible with semiconductor technology and allows to fabricate nanoscale devices (thickness of a
 conventional quarter-wave plate is much larger) with a simple operation (external magnetic field application is not easy).

 Recently Konishi et al.\cite{Konishi2011} fabricated a simple chiral nanostructure that consisted of  InAs quantum dots (QDs) embedded in the waveguide region
 of a planar GaAs based dielectric waveguide coupled to a dielectric chiral photonic crystal slab (CPCS), and obtained a degree of circular
 polarization of QD emission $\rho_c=26$\%.
A significantly higher value of $\rho_c$  of QD emission (up to 70\% in the direction
normal to the structure plane  and up to 81\% at some angle) was experimentally
achieved from a planar semiconductor microcavity with chiral partially etched top mirror,\cite{Maksimov2014} but this structure is
far more complex.
The degree of circular polarization of emission from a
chiral structure made of achiral semiconductors
depends strongly on its geometry and the radiation frequency, which allows one to optimize the
structures. Recent numerical simulations~\cite{Maksimov2014,Lobanov2015} show
that values of $\rho_c$ as high as 98 and 99\%  in waveguide and
microcavity structures, respectively, can be reached by their optimization.

In this paper, we have experimentally investigated the polarization of emission
of InAs QDs embedded in a planar GaAs
based dielectric waveguide coupled to a dielectric chiral photonic crystal slab with sub-wavelength period
in order to reach the predicted high values of $\rho_c$.
 The structure geometry
 has been optimized on the base of the computation of the QDs radiation performed using an oscillating point dipole model.\cite{Lobanov2012,Lobanov2015}
A circular polarization degree of InAs QD emission in the finite (NA less than 0.02) aperture as high as 81\% is experimentally realized in
 fabricated optimized structures. The achieved value is found to be limited by the unpolarized
scattered light background caused by the fluctuations in the
 periodic potential due to the size and shape imperfections of etched rectangular pillars in the chiral layer.
 It is worthwhile to note that the magnitude of $\rho_c$
 obtained from the spectra with canceled background due to the lattice imperfection reaches 96\%.
Thus, the considered very simple chiral structure made of conventional achiral semiconductor
materials using the chiral morphology effect can enable the fabrication of compact solid state circularly polarized light-emitting devices with
future progress in fabrication technology.

 \begin{figure}[h]
\includegraphics[width=0.8\linewidth]{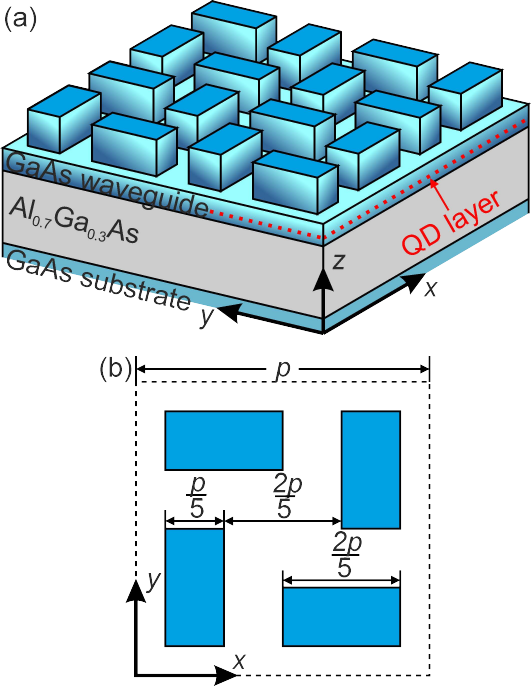}
\caption{\label{fig_01}
(Color online)
(a) Schematics of a structure with GaAs/air chiral photonic crystal slab, GaAs waveguide layer with an embedded layer of QDs,
and a Al$_{0.7}$Ga$_{0.3}$As cladding layer on GaAs substrate; (b) top view of one period  of the right
(or counterclockwise) twisted  structure. }
\end{figure}

In addition, we clarify the physical nature of the phenomenon.
Based on the reciprocity principle and the symmetry of the structure, we show the possibility to enhance and  filter out one circularly polarized
light component from the originally unpolarized QD emission. This is a consequence of the Fabry-Perot resonances between propagating photon modes
in the chiral photonic crystal slab. We show that the circular polarization efficiency of
such a chiral filter has to oscillate with its thickness, and the CPCS in the
optimized structure is actually a half-wave plate. Its thickness can be made however much less
than in the case of using natural optically active materials with a weak chirality,
of the order of the emitted light wavelength $\lambda$ only.
 It happens because in a dielectric CPCS with a high contrast of dielectric constants it becomes possible to achieve
a large difference between wavenumbers of different propagating modes, of the order of $k_0 = 2\pi/\lambda$.

We would like to add that recently, a possibility to control deterministically  the direction of photon emission (depending on the handedness of its circular polarization)
 by a QD inside a chiral photonic crystal waveguide has been demonstrated.\cite{Soellner2015} Our structure with randomly distributed QDs may open an alternative
 way  to this goal, being not sensitive to the exact positioning of the emitting QDs inside the structure.

The paper is organized as follows: The experimental details and the
	sample are described in  Sec.~\ref{Sec2} .  The experimental results are
	described  in Sec.~\ref{Sec3}. The physical nature of the effect is discussed in Sec.~\ref{Sec4}.  Section~\ref{Sec5} contains
	the concluding remarks.

\section{\label{Sec2} Experiment}
\begin{figure}[t]
\includegraphics[width=0.7\linewidth]{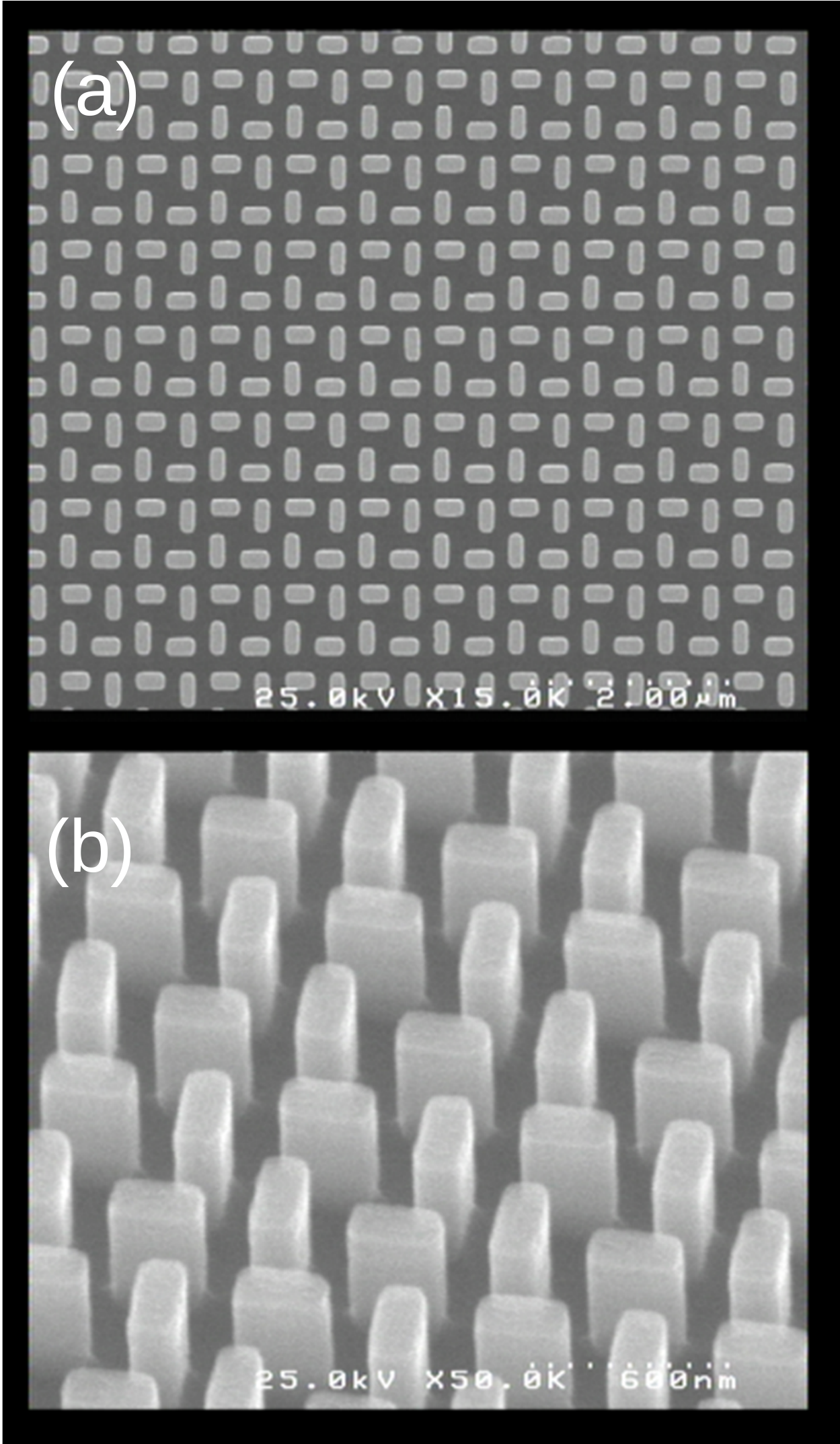}
\caption{ \label{fig_02}
Representative scanning electron microscopy (SEM) pictures of the right-twisted chiral photonic
crystal fabricated from the GaAs-based waveguide, top (a) and side (b) view. }
\end{figure}

The epitaxial layer structure, which formed the basis of this study, consisted of a
691~nm thick GaAs membrane with a single layer of embedded randomly distributed InAs quantum dots (Fig.~\ref{fig_01}a).
The membrane is grown on top of a 1~$\mu$m thick AlGaAs buffer layer.
We fabricated an array of nano-pillars into the upper nominally 428~nm thick part of the membrane layer via electron beam lithography
 and reactive ion etching. The nano-pillars were arranged into
a square lattice of unit cells consisting of four
rectangular pillars, each rotated by 90$^\circ$ with respect to its nearest neighbors. The
lattice period $p$ was varied in the range of 0.7--1.1\,$\mu$m. The unit cell is
shown in Fig.~\ref{fig_01}b. The distance between pillars was $p/5$, their sides were $p/5$
and $2p/5$, respectively. Both right  and left (mirrored)  twisted pillars were manufactured.
(For definiteness, we call hereinafter the structures shown in Figs.~\ref{fig_01},\ref{fig_02} as right-twisted.)
The thickness of GaAs waveguide layer ($\epsilon_\mathrm{GaAs} =
12.42$), containing a plane with QDs was 263~nm. The plane with QDs was located 100~nm above its bottom. The
lower Al$_{0.7}$Ga$_{0.3}$As ($\epsilon_\mathrm{AlGaAs}= 9.66$) buffer layer was 1 $\mu$m thick. The whole
system was grown by molecular beam epitaxy on a GaAs substrate. Representative
SEM pictures of the right-twisted chiral photonic layer are shown in Fig.~\ref{fig_02}.

The sample was at a temperature of 5~K in the insert of an optical cryostat. The luminescence
polarization was analyzed by a quarter wave retarder and linear polarizers. The angle resolution of
0.5$^\circ$ is ensured by a small aperture. The circular polarization degree of the luminescence
from the structure prior to etching was less than 0.1\%.\\
\begin{figure}[h]
\includegraphics[width=0.8\linewidth]{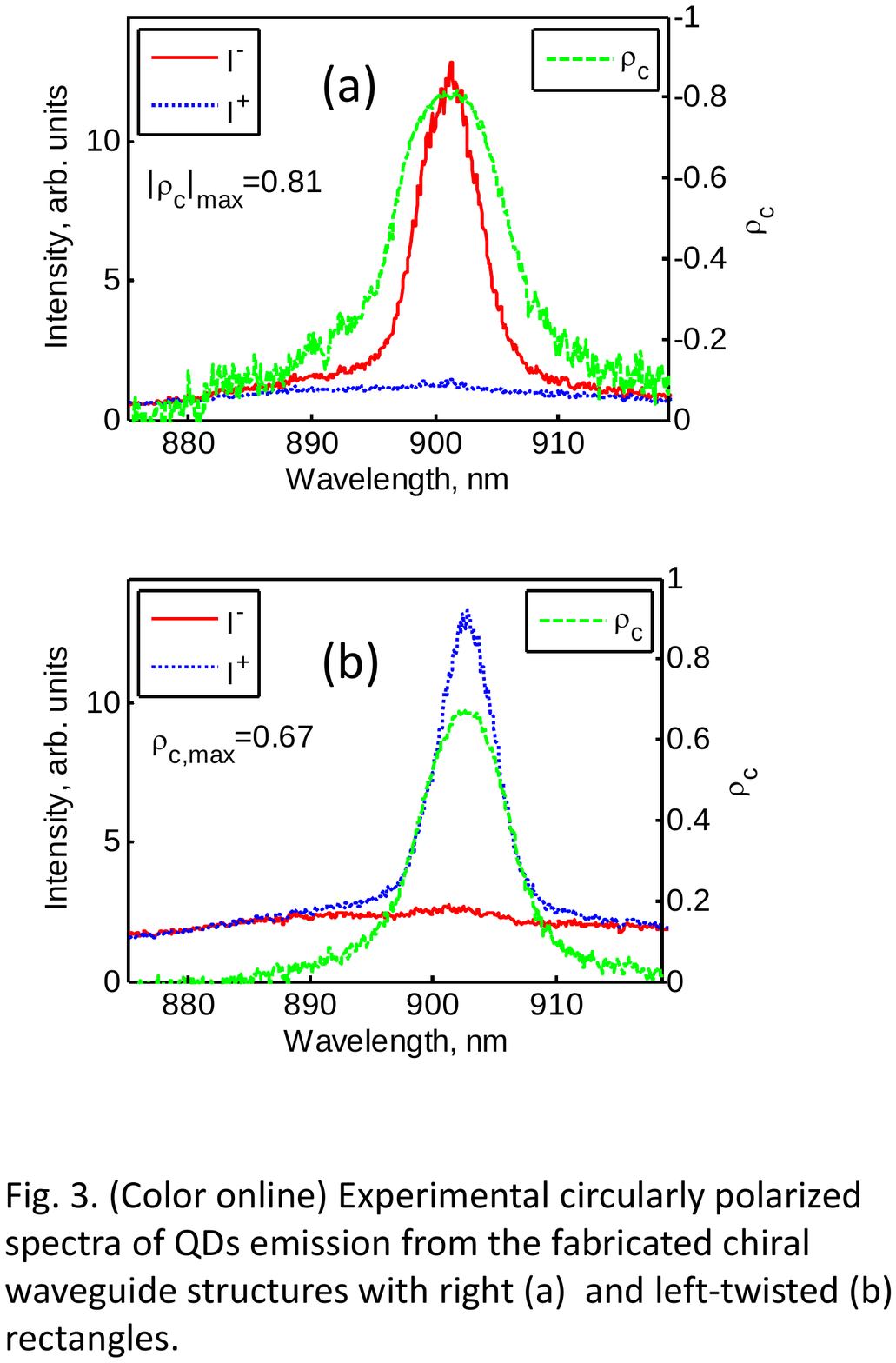}
\caption{ \label{fig_03}
 (Color online)
Experimental circularly polarized spectra of QD emission from the fabricated chiral waveguide
structures with  (a) the right-  and (b)  the left-twisted rectangular pillars. Green dashed lines show the corresponding
circular polarization degree  $\rho_c$.}
\end{figure}

\begin{figure}[b]
\includegraphics[width=0.8\linewidth]{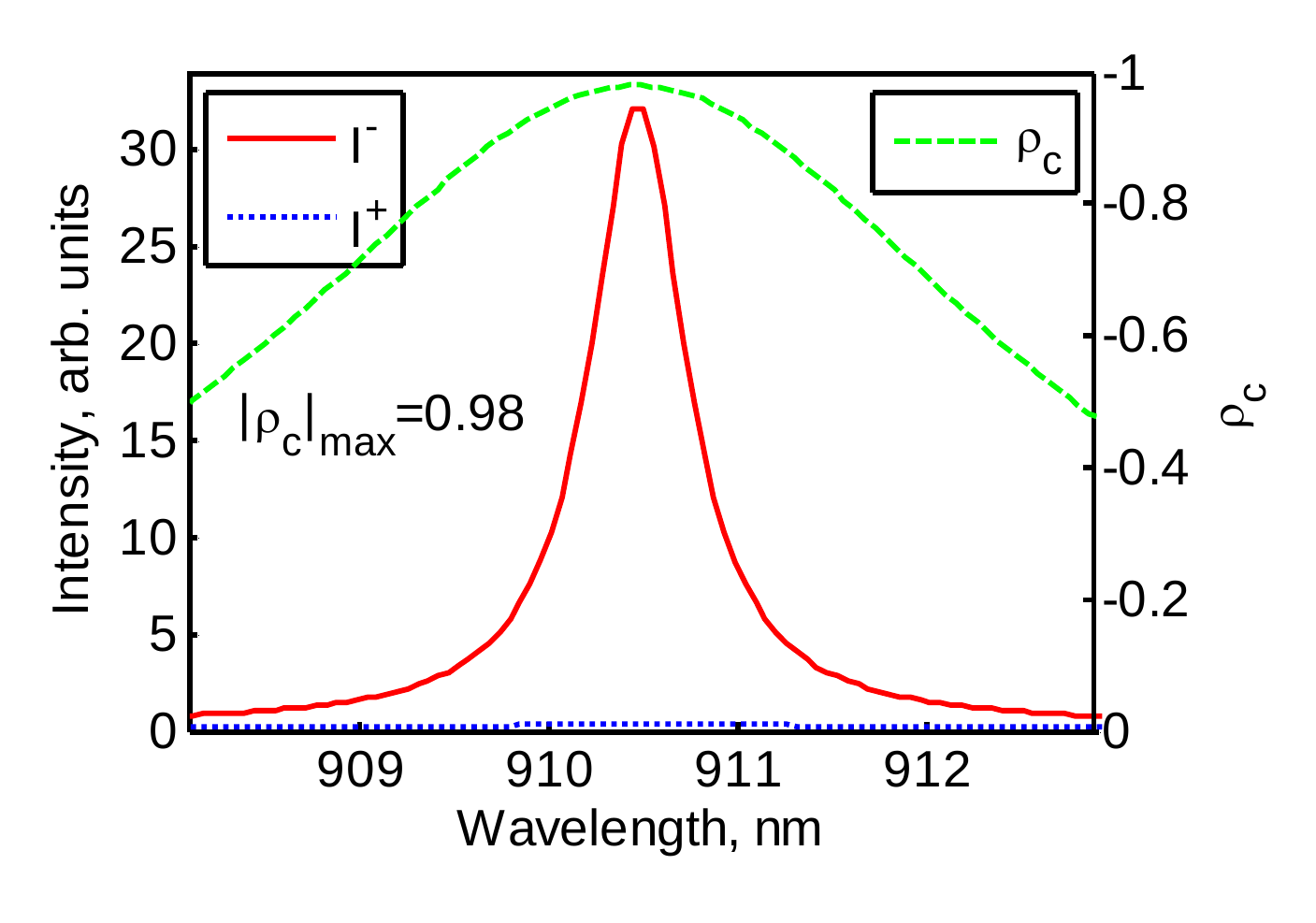}
\caption{ \label{fig_04}
 (Color online) Calculated circularly polarized spectra and circular polarization degree of QD emission
 from the  chiral waveguide structure with the right-twisted pillars. The sign of circular
  polarization is opposite for the left-twisted pillars. The emission intensity is normalized to the emission intensity of the
  same oscillating dipole in vacuum (without the structure). }
\end{figure}

\begin{figure}[h]
\includegraphics[width=0.8\linewidth]{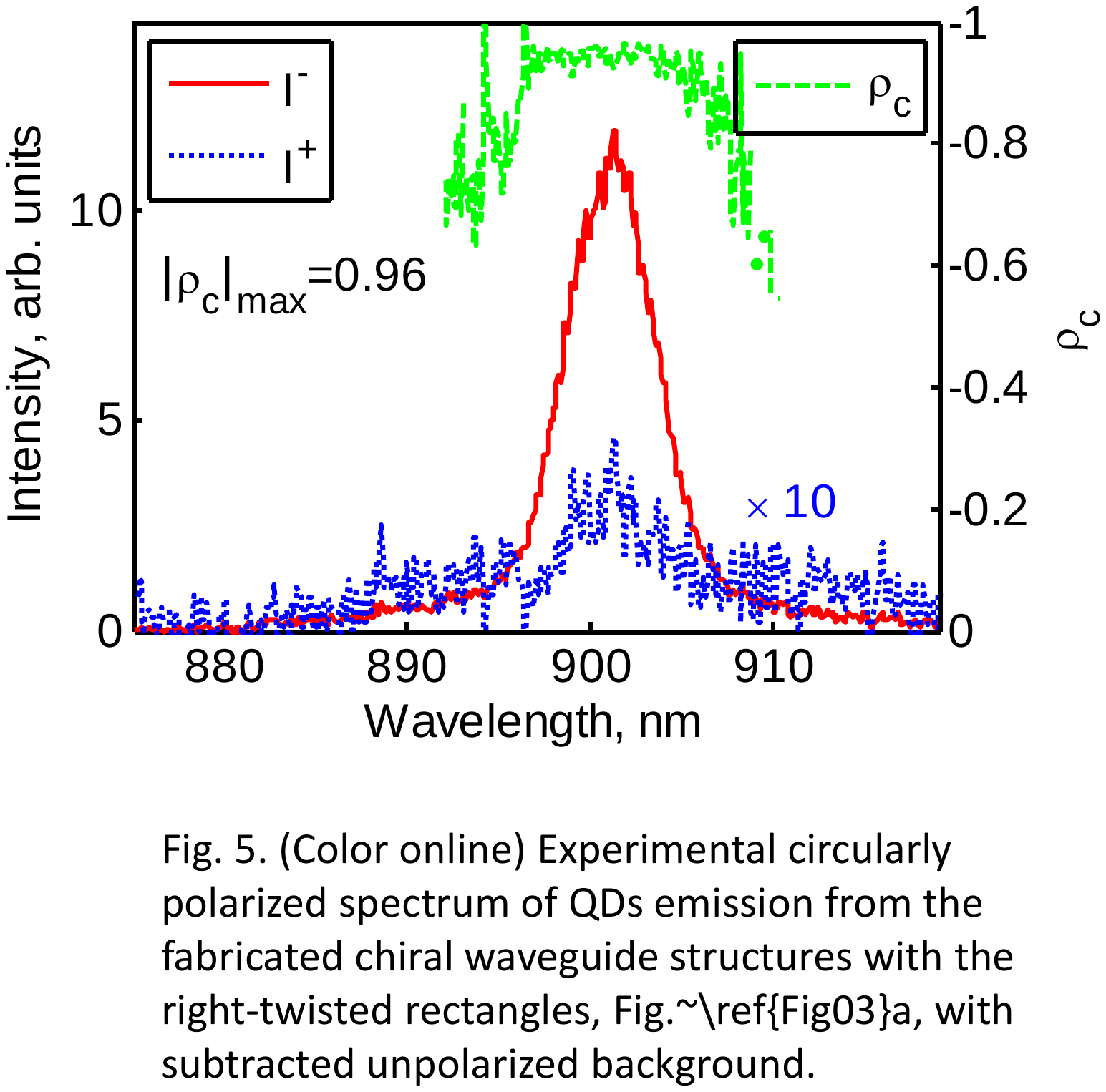}
\caption{ \label{fig_05}
 (Color online)
Experimental circularly polarized spectrum of QD emission from the fabricated chiral wave\-guide
structures with the right-twisted pillars, Fig.~\ref{fig_03}a, with subtracted unpolarized background.}
\end{figure}

\begin{figure}[h]
\includegraphics[width=0.9\linewidth]{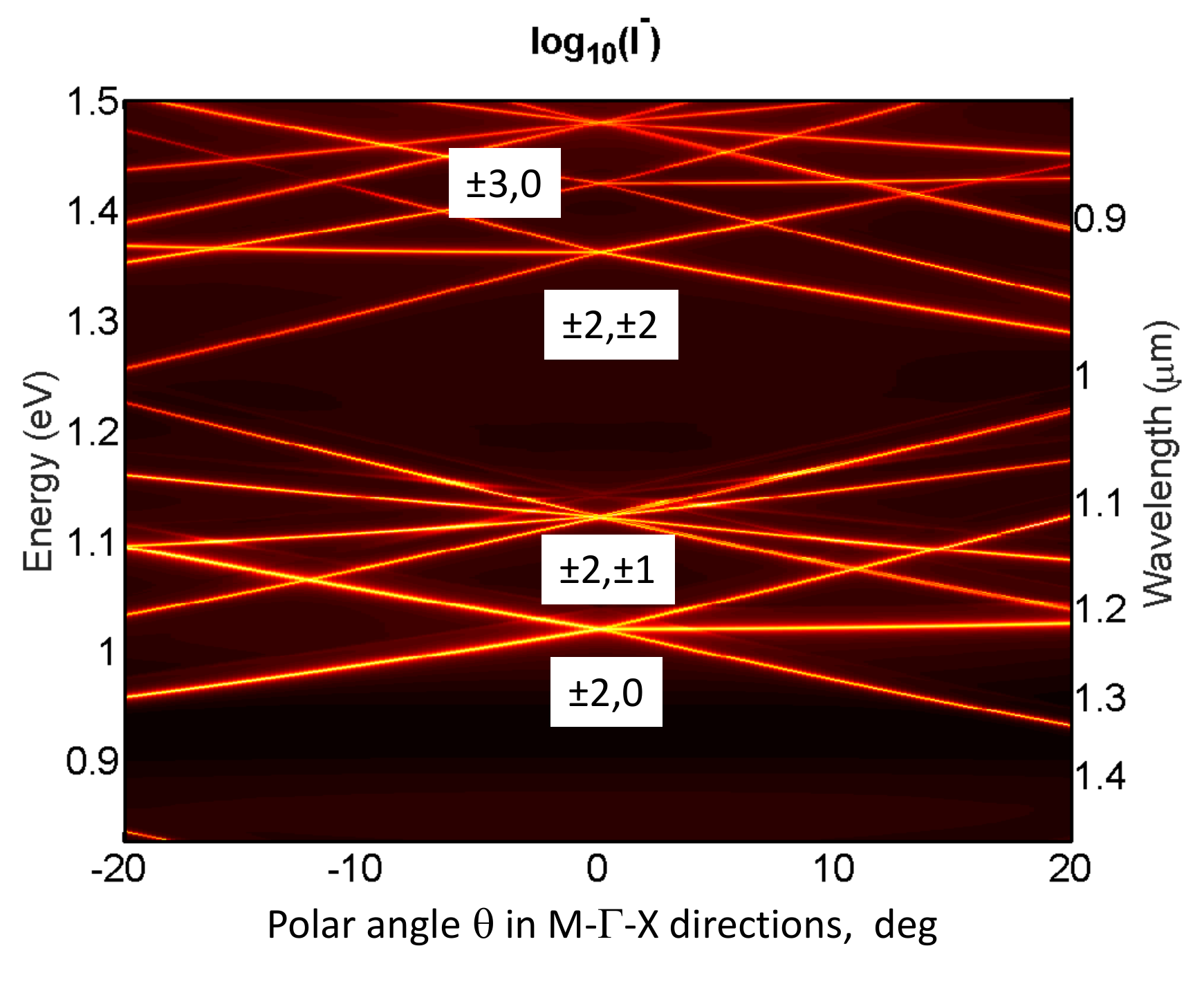}
\caption{ \label{fig_06}
(Color online) Dispersion of quasiguided modes in spatially structured GaAs waveguide. False color scale shows the calculated QD
emission intensities (in left polarization, logarithmic scale) at various $k = k_0 \sin \theta$, $k_0 = \omega/c$, along
$\Gamma – X$ and $\Gamma – M$ (diagonal) directions in the 1st BZ.  Labels $l,m$ (see in the text) mark the order of the
corresponding Bragg resonance in the $\Gamma$-point. }
\end{figure}

 \begin{figure*}[t]
\includegraphics[width=0.45\linewidth]{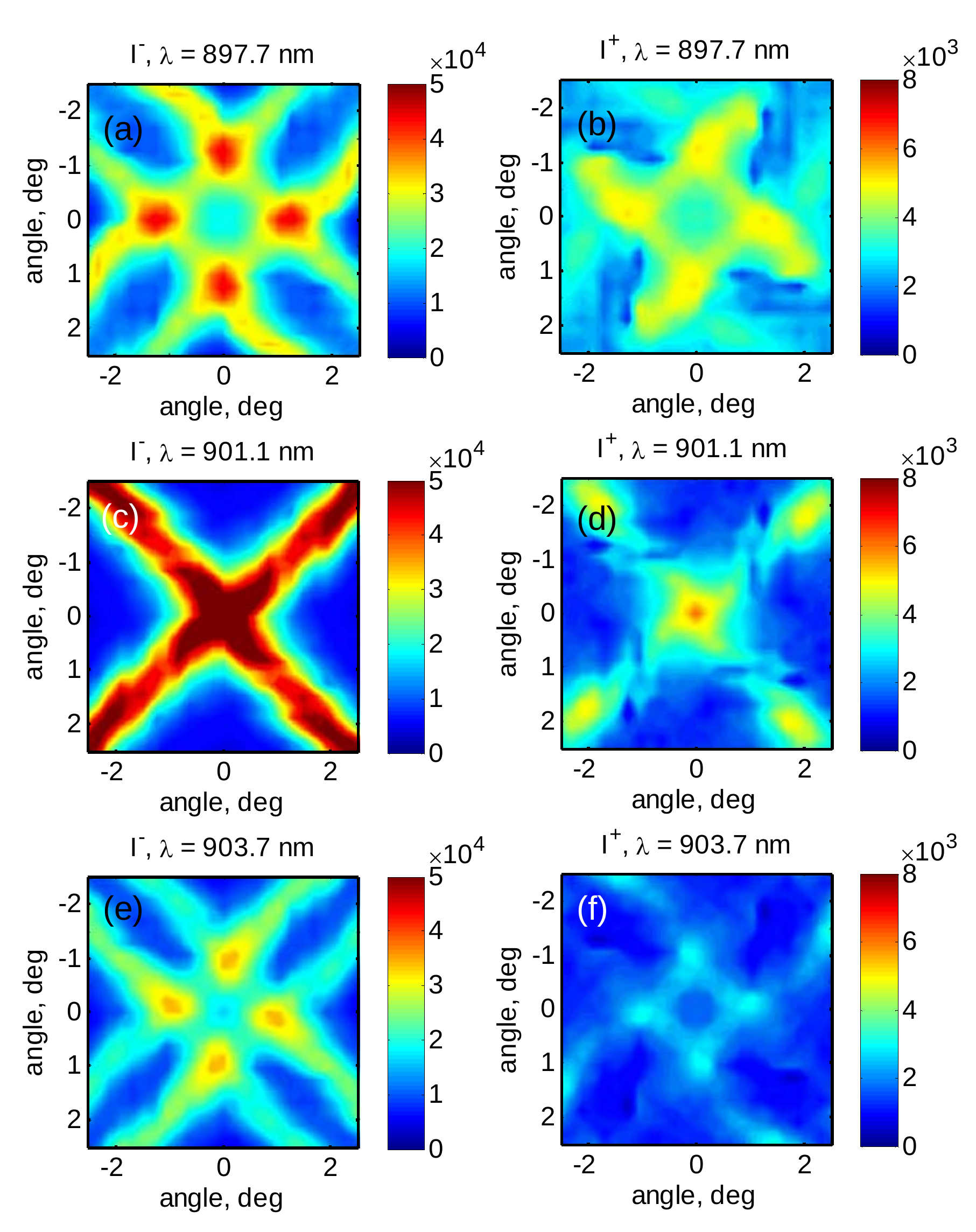}
\includegraphics[width=0.45\linewidth]{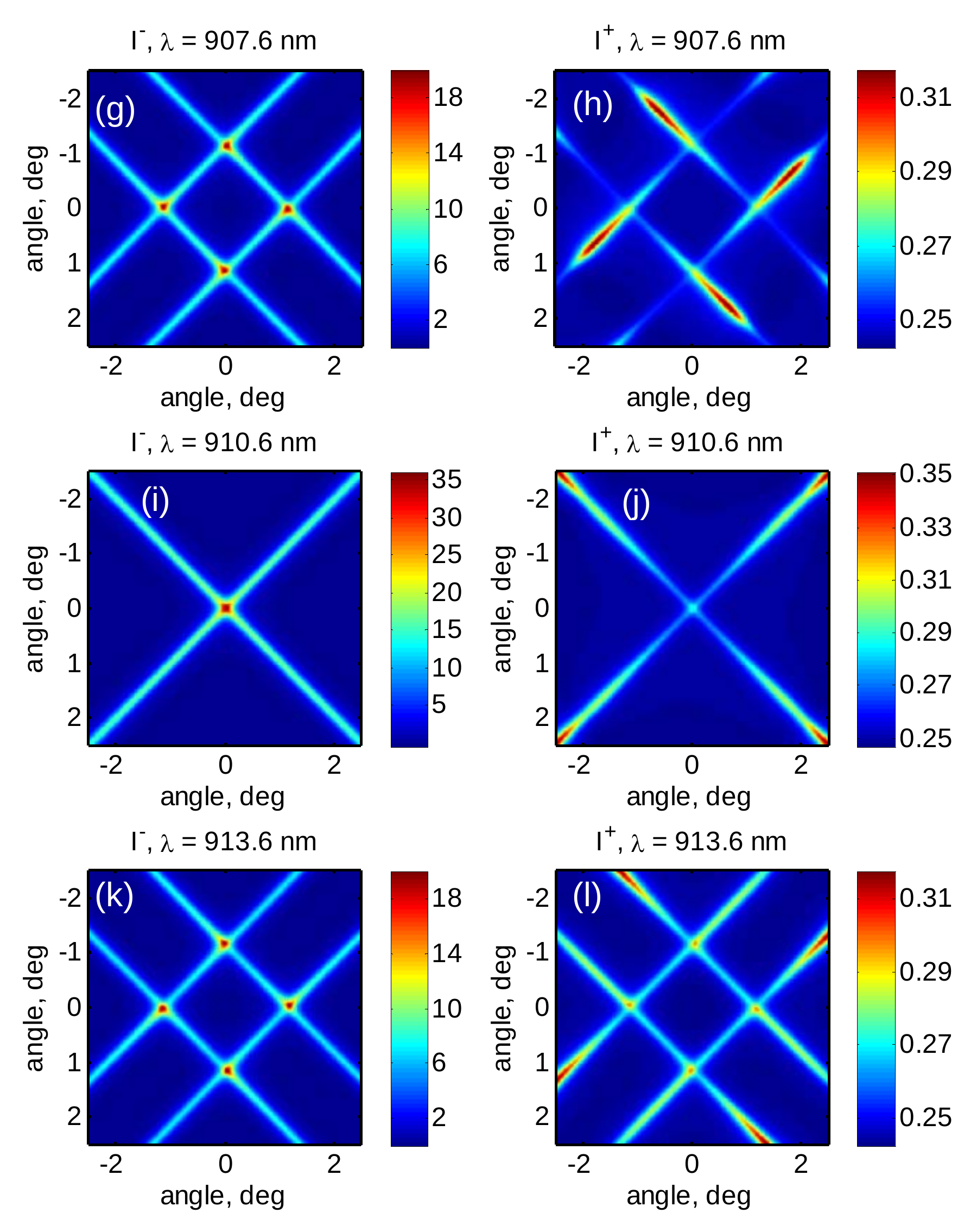}
\caption{ \label{fig_07}
(Color online)  Polar plots of the measured (left panels) and calculated (right panels) angle dependencies of the QD emission intensities in left and
right circular polarizations. The measured spectra were recorded for the resonance wavelength $\lambda_0 = 901.1$~nm (panels c and d),
corresponding to the ($\pm 2, \pm2$)  Bragg resonance at $E=E_0$ (see in the text). Panels (a,b) and (e,f) show the corresponding
dependencies for $\lambda_0-3.4$ and $+2.6$~nm. The simulated spectra were calculated around  the theoretical resonance wavelength
$\lambda_{0,\mathrm{th}} = 910.5$~nm (panels i and j). Panels (g,h) and (k,l) show the corresponding
dependencies for $\lambda_{0,\mathrm{th}} \mp 3$~nm.
  }
\end{figure*}

\section{\label{Sec3} Experimental results}

Experimental circularly polarized spectra of QD emission are shown in Fig.~\ref{fig_03} (left axis) for
left and right-twisted rectangular pillars.
The green lines in Fig.~\ref{fig_03}, right axes,  show the resulting  spectra of the circular polarization degree
of emission
\begin{equation}\label{rhoc}
    \rho_c=\frac{I^+-I^-}{I^++I^-},
\end{equation}
where
$I^{\pm}$ is the emission intensity in right/left circular $\sigma^\pm$-polarization.\footnote{The sign of circular
polarization is defined from the point of view
of the receiver. The electric field of a planar left ($\sigma^-$) circularly polarized wave, propagating in $z$-direction
and rotating counter-clockwise from the point of view of the receiver is
$E^-\propto (\mathbf{e}_x+i\mathbf{e}_y)\exp(ik_0z-i\omega t)$.}
The spectra are recorded in the direction $z$ (perpendicular to the structure plane)
with numerical aperture NA=0.005.
For comparison, the calculated emission spectra in $\sigma^\pm$ polarization and $\rho_c$
are shown in Fig.~\ref{fig_04} for the right-twisted structure with the same parameters.
The spectra were calculated using  the reciprocity principle and the optical scattering
matrix method.~\cite{Maksimov2014,Lobanov2015} Compared to Ref.~\onlinecite{Lobanov2015}, we
changed the parameters of the materials, because the current experiment is carried out at low  rather
than at room temperature.

Figure~\ref{fig_03} shows that, in accordance with the
calculations, the polarization of the emission line at the resonance
frequency is opposite in the structures with the left and right-twisted pillars.

The experimental and calculated dependencies in Figs.~\ref{fig_03} and \ref{fig_04} are in a good qualitative agreement.
A small 9~nm red shift of the
calculated resonance wavelength is most probably due to a slightly incorrect dielectric susceptibilities used in the calculations, the agreement
can be easily improved by their fitting.

Note, however, that (i) the experimental line FWHM is
about 5 times greater than the calculated one ($\approx 4$ and 0.8 nm, respectively),
and (ii) the line sits on a background whose magnitude is
about 9\% from its peak intensity whereas the background in the calculated spectra does not exceed 1\%.
 Both the line broadening and background magnitude are
found to depend strongly on the pillar quality, which indicates small deviations from the ideal structure.
Figure~\ref{fig_03}  shows that magnitudes of
$\rho_c = - 0.81$ and +0.67 in the waveguides with the right and left-twisted pillars
are limited by the unpolarized background. It is worthwhile to note that the magnitude of $|\rho_c|$
at the resonance reaches 0.96 when this unpolarized background is removed as shown in Fig.~\ref{fig_05}.
This value is very close to the calculated one,
which indicates that the predicted values of $|\rho_c| \approx 0.98$ could be reached due to further
advances in fabrication technology.

\section{\label{Sec4} Discussion}
\subsection{\label{Subsec4.1} Angular dependencies of emission intensities, $\rho_c$, and the quasiguided
modes}

As already discussed,~\cite{Konishi2011,Lobanov2015}  the peaks in the QD emission spectrum
are due to the excitation of lossy  quasiguided modes in the GaAs waveguide inside a periodically modulated
structure. Their expected angular dispersion is shown in  Fig.~\ref{fig_06}, where false colors show the
calculated intensity of the emission (in $\sigma^-$ polarization) from the right-twisted
structure with period $p=770$~nm as a function of energy (wavelength) and polar angle $\theta$ in
 $M-\Gamma$ (diagonal)  and $\Gamma-X$ directions.
These Bragg resonances can be understood~\cite{Tikhodeev2002}  as corresponding approximately to the
folding of the dispersion of the guided modes into the 1st Brillouin zone (BZ) of the 2D grating (square lattice  in our case),
\begin{equation}
\label{Eq1}
E_{l,m}(\mathbf{k})=\frac{\hbar c}{n_\mathrm{eff}} \sqrt{(k_x+lK_\mathrm{BZ})^2 +(k_y+mK_\mathrm{BZ})^2},
\end{equation}
where $\mathbf{k} = (k_x,k_y)$ is the in-plane wavevector,  $K_\mathrm{BZ}=2\pi / p$ is the Bragg wavenumber,
$p$ is the period, $n_\mathrm{eff}$ is the effective refractive index of the guided mode,
$l$ and $m$ are integers.
It appears that the resonance of interest at $\lambda_0 \sim 901$~nm in the current structure
corresponds to  a fourfold degenerate $l,m=\pm2$ state  in the $\Gamma$-point (in the scalar approximation and neglecting the energy
splitting due to the point symmetry group), see Fig.~\ref{fig_06}.
It is also in agreement with the analysis of Ref.~\onlinecite{Lobanov2015},
see, e.g., the calculated in-plane distribution of the emission intensity in the dominant polarization in Fig. 6b of Ref.~\onlinecite{Lobanov2015}.

 \begin{figure}[t]
\includegraphics[width=0.98\linewidth]{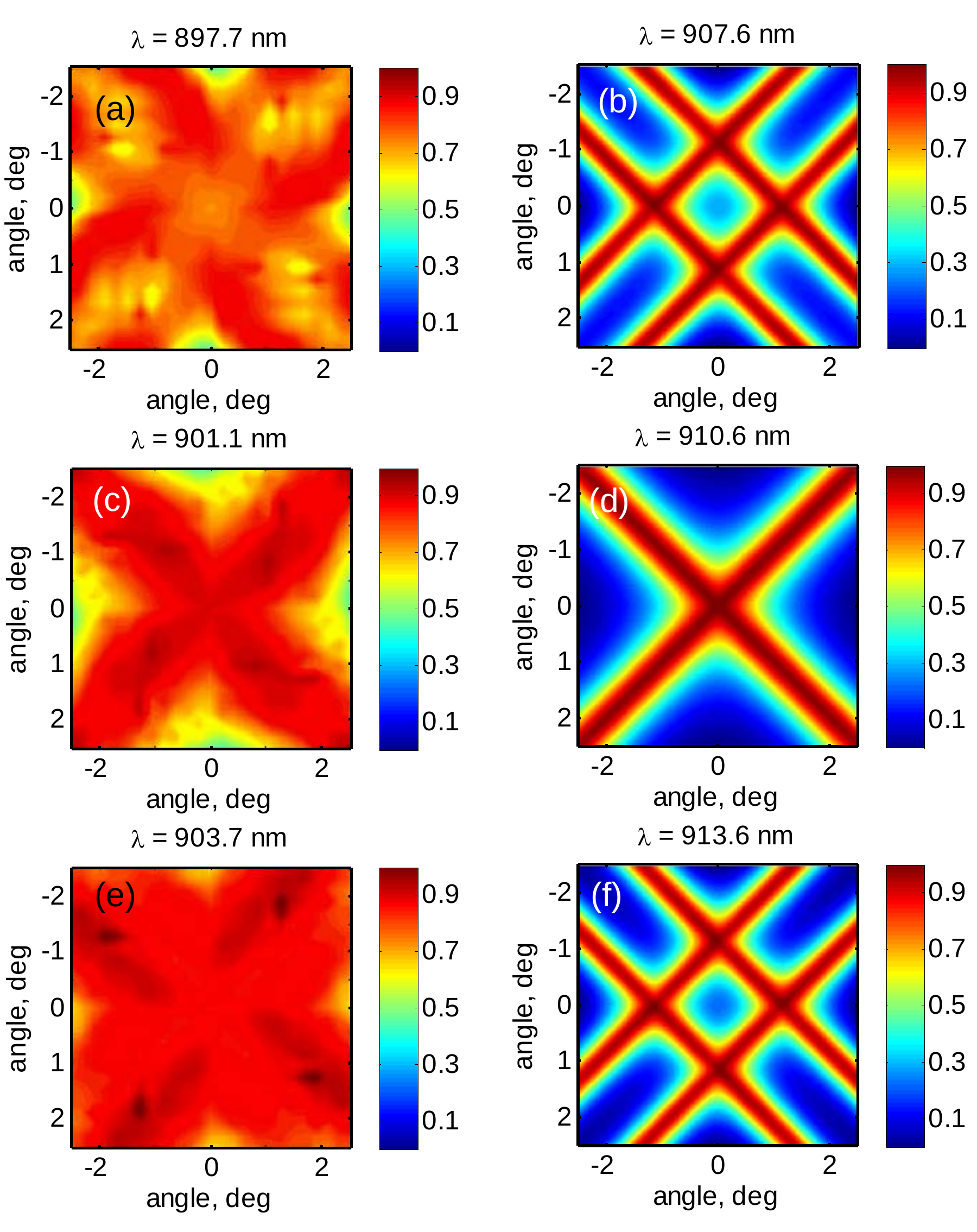}
\caption{ \label{fig_08}
(Color online) Polar plots of the measured (left panels) and calculated (right panels) angle dependencies of $|\rho_c|$ of the QDs emission,
corresponding  to the data in Fig.~\ref{fig_07}.}
\end{figure}

The degeneracy of the state $E_{\pm2,\pm2}$ is lifted at $k\neq 0$.
According to Eq.~\ref{Eq1}, near the $\Gamma$-point
\begin{equation}\label{Gamma}
    E_{l,m}(\mathbf{k})= E_0+ \frac{\hbar c (lk_x+mk_y) }{n_\mathrm{eff} \sqrt{l^2+m^2}}+\mathcal{O}(k_x^2+k_y^2),
\end{equation}
where $E_0$ is the resonance energy in the $\Gamma$-point, $E_0 \equiv E_{\pm2,\pm2}(k=0)$.
Thus, in the $0-X$, $0-Y$ directions, according to Eq.~\ref{Gamma}, the ($\pm 2, \pm 2)$ modes split into
two degenerate doublets with nearly linear $k$-dispersion
$E-E_0=\pm \hbar c k_{x,y} /(\sqrt{2}n_\mathrm{eff})$.
 In the diagonal directions,  the modes with $l= m=\pm 2$ split into  two
singlets with nearly linear dispersion $E-E_0=\pm \hbar c k /n_\mathrm{eff}$,
whereas those with $l=-m=\pm 2$  is a degenerate  parabolic
doublet with an effective mass $m_\mathrm{eff}\sim 10^{-5} m_e$, where $m_e$ is
the free electron mass.
Thus, the modes with $l=-m=\pm 2$
are nearly dispersionless at small $k\lesssim 0.2$~$\mu$m$^{-1}$.
As a result, the lines of equal energy at $E=E_0$  are
two mutually perpendicular lines  at $k_y=\pm k_x$ in this range of $k$.
That is well seen in Figs.\ref{fig_07}c,d representing the measured  angular
distribution of the QD emission near $\theta=0$ in two circular
polarizations at the resonance wavelength.

Figures~\ref{fig_07}a,b and e,f show that these lines
at  $E\neq E_0$  split into doublets  $k_y=k_x\pm \Delta k$  and
$k_y=-k_x\pm \Delta k$ in accordance with the $D_4$ point group symmetry
of the modes' dispersion at the $\Gamma$ point.

 The measured anglular distributions of the emission intensities in $\sigma^\pm$ polarizations and of the circular polarization degree (left panels
 of Figs.~\ref{fig_07} and ~\ref{fig_08} respectively)
  are in a good qualitative agreement with the calculated ones displayed in the right panels. It is seen in
  Fig.~\ref{fig_07} that, in agreement with the calculation,  the strongest emission in the dominated ($\sigma^-$) polarization is
  at the crossing of the  isoenergetic lines and
  reaches its maximum at $E=E_0$ when the state at $k=0$ becomes fourfold degenerate. According to calculations
  shown in Fig.~\ref{fig_04} the QD emission intensity in the dominating left polarization at the resonance  $E=E_0$
   exceeds that in vacuum more than 30 times, whereas
  that in the opposite  polarization  is strongly (about 3 times) suppressed. The effect is connected to a modification of the
  environmentally allowed electromagnetic modes by the CPCS in $k$-space. As a consequence, it is very sensitive to the
  quality of the photonic crystal. Any disturbance of the potential periodicity results in the increased Rayleigh scattering
  contributing to $k=0$ emission in the suppressed polarization.  The result is  the decreased degree of circular
   polarization in the experiment.

\begin{figure}[t]
\includegraphics[width=0.99\linewidth]{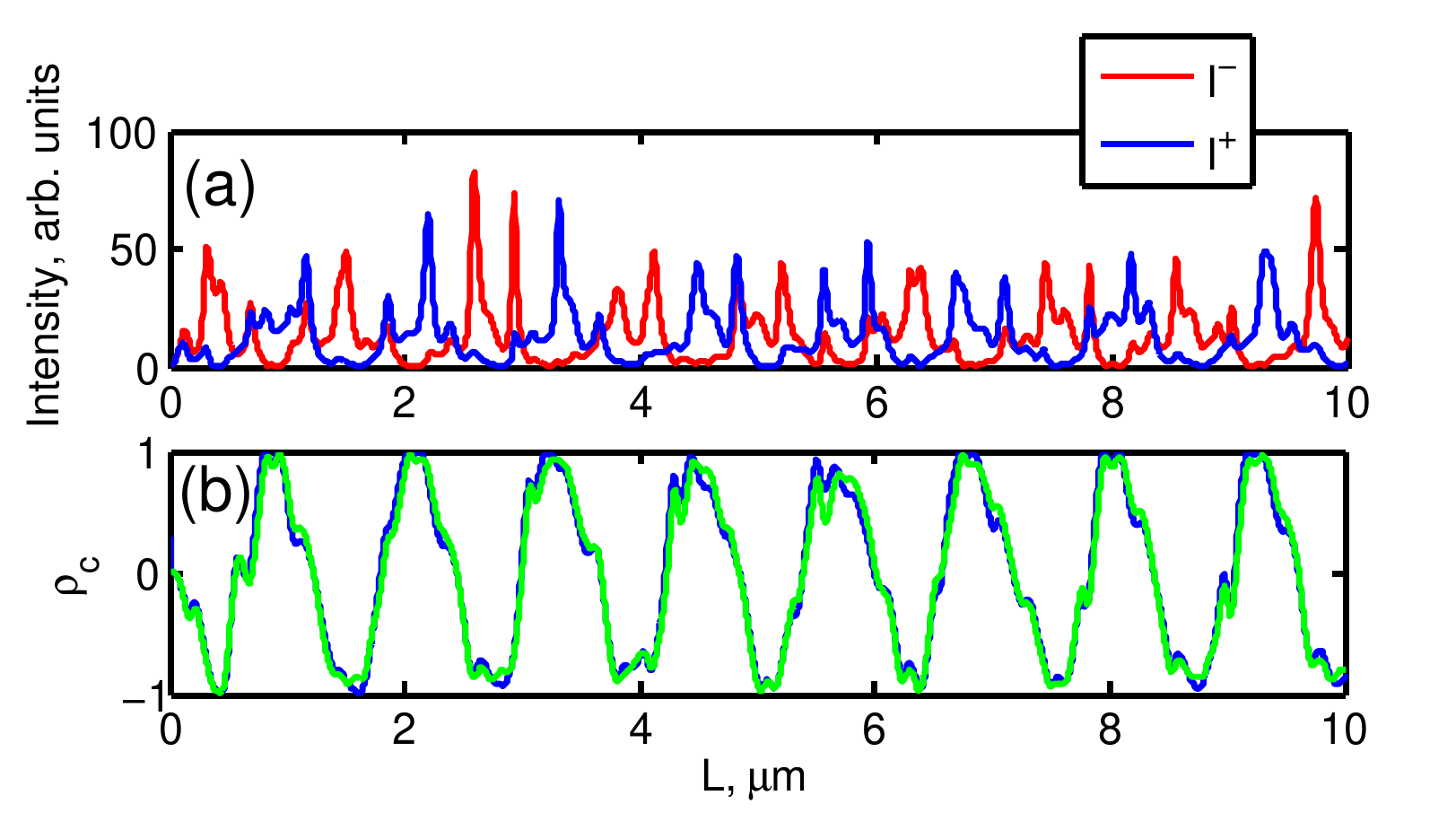}
\caption{ \label{fig_09} (Color online)  The calculated dependencies of (a)  the  emission intensities  $I^\pm$  and (b, green line) the circular polarization degree
of emission $\rho_c$ at the resonance wavelength $\lambda_{0,\mathrm{th}}=910.5$~nm on the chiral photonic crystal slab  thickness $L$. Blue line in panel  (b) shows
the calculated circular selectivity of  the chiral slab  transmission $\rho_{c,T} = (T^+-T^-)/(T^++T^-)$, see the discussion in Subsec.~\ref{Subsec4.2}.
}
\end{figure}

\subsection{\label{Subsec4.2} Symmetry analysis and the physical origin of circularly polarized emission}

Although  the possibility to obtain  nearly 100\% circularly polarized emission from unpolarized emission of randomly
distributed  QD  was already demonstrated numerically,~\cite{Maksimov2014,Lobanov2015} the physical
mechanism of this effect has not been yet clarified.
The key property of the structure that helps to understand this mechanism is the expected theoretically oscillatory dependence of the degree of circular
polarization of emission of such structures on the thickness of the chiral photonic slab thickness, illustrated in Fig.~\ref{fig_09}.
While the calculated dependencies of the emission intensities in right and left polarization $I^\pm$ on $L$ (see Fig.~\ref{fig_09}a)
consist  of series of at first sight incoherent peaks, the dependence $\rho_c(L)$ (see green line in Fig.~\ref{fig_09}b) looks as coherent Fabry-Perot oscillations.
 In what follows, based on the reciprocity and symmetry properties
of our structure, we show that the chiral photonic crystal slab in the structure of interest plays the role of a wave plate.
 It introduces a phase shift, proportional to $\Delta K_z L$ ($L$ is the CPCS thickness),  between the propagating modes
with a large (of the order of $k_0 = \omega/c$) difference of wavenumbers $\Delta K_z $.
We will show that these modes are basically different waveguide modes in the GaAs rectangular vertical waveguides, i.e., in the
building blocks of the CPCS.
There is also a large constant phase shift between left and right circular polarizations, due to the absence of horizontal
mirror symmetry of the photonic crystal slab with different top and bottom boundaries (air/CPCS and CPCS/GaAs, respectively).
As a result,   the  transmission of the CPCS in  left and right polarizations oscillate as a function of $L$ in
counter-phase, which opens a way to reach nearly 100\% polarization selectivity.

\subsubsection{\label{Subsubsec4.2.1} Reciprocity principle for the structures with C$_4$    and D$_4$ symmetry}
The emission intensity of QDs can be calculated directly~\cite{Whittaker1999,Whittaker2000,Taniyama2008,Lobanov2012}
or using the electromagnetic reciprocity principle.\cite{LLECM1984,Maksimov2014}
 In the latter case, the emission problem of randomly distributed and randomly polarized oscillating dipoles is replaced
by an illumination problem, in which the structure is illuminated by a polarized plane wave
$\mathbf{E}_\mathrm{in} (x,y,z)=E\mathbf{e}^{j} \exp(i k_0 z-i\omega t)$.
(Hereinafter index $j$ marks the polarization state, e.g., $j=x,y$  in case of linear polarization along $x,y$, respectively,
and $j=\pm$ for right and left polarization.)
Then, the resulting electric field distribution in the layer with QD,
\begin{equation}\label{Int}
    I^{j}=\iint \mathrm{d}x\mathrm{d}y \left(|E_x^j |^2+|E_y^j |^2\right) ,
\end{equation}
gives the corresponding polarized component of the QD radiation intensity.\cite{Maksimov2014}

The reciprocity principle helps to understand  easily, why in the structure with  a C$_4$
symmetry, the emission of QD in the vertical direction cannot be linearly polarized, and why for a structure
with higher symmetry (e.g., D$_4$), it cannot be also circularly polarized.

Assume  $\mathbf{E}(x,y,z)$ and $\mathbf{H}(x,y,z)$  are the solutions of Maxwell's
equations for the system with C$_4$ symmetry. Then, the electromagnetic field rotated by 90$^\circ$  is also a solution.
The 90$^\circ$ rotation switches the linear polarization of the illuminated plane wave $j=x$
to the orthogonal polarization $j=y$,
while the integral (\ref{Int}) does not change, $ I^x = I^y$. Because of the Eq.~\ref{Int},
$\rho_{xy} =( I^x - I^y)/( I^x + I^y)=0$, hence the emission cannot be linearly polarized.

Similarly, if a structure has a
D$_4$ symmetry (equivalent to C$_{4v}$), Maxwell's equations are invariant also under mirror reflection
operation, e.g.,  in $XOZ$ plane, which
switches the sense of circular polarization of the illuminated plane wave
$j=\pm$ to the opposite one,
$j=\mp$, while the integral
(\ref{Int}) does not change, $ I^+ = I^-$.  Thus, again, the degree of circular polarization of emission
$\rho_c =( I^+ - I^-)/( I^+ + I^-)=0$, hence the emission cannot be circularly polarized, too.


\subsubsection{\label{Subsubsec4.2.2} S-matrix approach and C$_4$ symmetry}
In the structure of interest with C$_4$ symmetry, a layer with randomly distributed and randomly polarized QD is placed in the planar GaAs waveguide.
Two lower effective dielectric permittivity layers operate as mirrors. The bottom mirror is the Al$_{0.7}$Ga$_{0.3}$As layer.
The top mirror is the 2D chiral photonic crystal.
The propagation of light in this system can be described by the optical scattering matrix. The basic idea
 is (i) to split the system into several layers in which the dielectric permittivity does not depend on $z$, and (ii) to find the 2D electromagnetic
modes of each slab, their electromagnetic field distributions $\left\{ E_x (x,y),E_y (x,y)\right\}$,  $\left\{H_x (x,y),H_y (x,y)\right\}$, and
 wavenumbers $K_z$. The periodicity of the structure along $x$ and $y$ axes leads the
fields to be  represented as  products of  periodic functions and harmonics $\exp(i(k_xx+k_yy))$, where $k_{x,y}$ are in the
1st BZ, with $-K_\mathrm{BZ}/2 < k_{x,y}<K_\mathrm{BZ}/2$.

The C$_4$ symmetry leads the in-plane components of the electric and magnetic fields to be
represented after a 90$^\circ$ rotation in one of the four forms
\begin{equation}\label{Fxy}
\left\{F_x (-y,x),F_y (-y,x)\right\}=i^n \left\{F_y (x,y),-F_x (x,y)\right\},
\end{equation}
where $n =  -1,0,1,2$, and  $F$ means the electric or magnetic fields. The advantages of the periodicity and C$_4$ symmetry can be
simultaneously used only for $k_x=k_y = 0$. In this case, the left polarized light
$\mathbf{E}^-= E \frac{\mathbf{e}_x+i\mathbf{e}_y}{\sqrt{2}} \mathrm{e}^{i k_0 z-i\omega t}$,
propagating in air normal to the structure plane, corresponds to $n = -1$, while the right polarized one
corresponds to $n = +1$. Since two other forms are optically inactive (see, e.g., in Ref.~\onlinecite{Tikhodeev2002}),
we will consider later only $n = \pm 1$.\\

\subsubsection{\label{Subsec4.2.3} Time-reversal symmetry}

The time-reversal symmetry of Maxwell's equations for systems consisted of non-absorbing and non-gyrotropic materials relates
wavenumbers and electromagnetic fields of the propagating ($\mathrm{Im} K_z=0$) modes with opposite $n$. If $\mathbf{E}_{3D} (x,y,z)$ and
$\mathbf{H}_{3D} (x,y,z)$ are the solutions of Maxwell's equations for infinitely thick layer, the time-reversal symmetry requires $\mathbf{E}^*_{3D} (x,y,z)$ and
$-\mathbf{H}^*_{3D} (x,y,z)$  to be the solutions, too.

Considering the propagation of some mode with fields $\left\{E_x (x,y),E_y (x,y)\right\}\mathrm{e}^{iK_z z}$ and
$\left\{H_x (x,y),H_y (x,y)\right\}\mathrm{e}^{iK_z z}$, one obtains that $\left\{E^*_x (x,y),E^*_y (x,y)\right\}\mathrm{e}^{-iK_z z}$ and
$-\left\{H^*_x (x,y),H^*_y (x,y)\right\}\mathrm{e}^{-iK_z z}$ are also solutions. It corresponds to the counter propagating mode with opposite
polarization $n$, since complex conjugation of one polarization ($n=+1$ or $-1$) switches the polarization to the opposite  one ($n=-1$ or $+1$, respectively).
As a result,
\begin{equation}\label{pm}
K_z^+=K_z^-, E^+=(E^- )^*,  H^+=(H^- )^*,
\end{equation}
where the upper indices $\pm$ show the values of $n$.
Note that the eigenmodes are always defined up to an arbitrary factor $C$. If propagating eigenmodes are normalized to have fixed energy flow
(we choose $\frac{c}{8\pi}$ per elementary unit cell), the eigenmodes are still defined up to an arbitrary factor with unitary absolute value $\mathrm{e}^{i\psi}$.
Hereinafter, we choose the phase $\psi$ for the opposite circular polarizations so that their electromagnetic fields are complex
conjugate of each other.

\subsubsection{\label{Subsubsec4.2.3} Unitary scattering matrix}
\begin{figure}[h]
\includegraphics[width=0.5\linewidth]{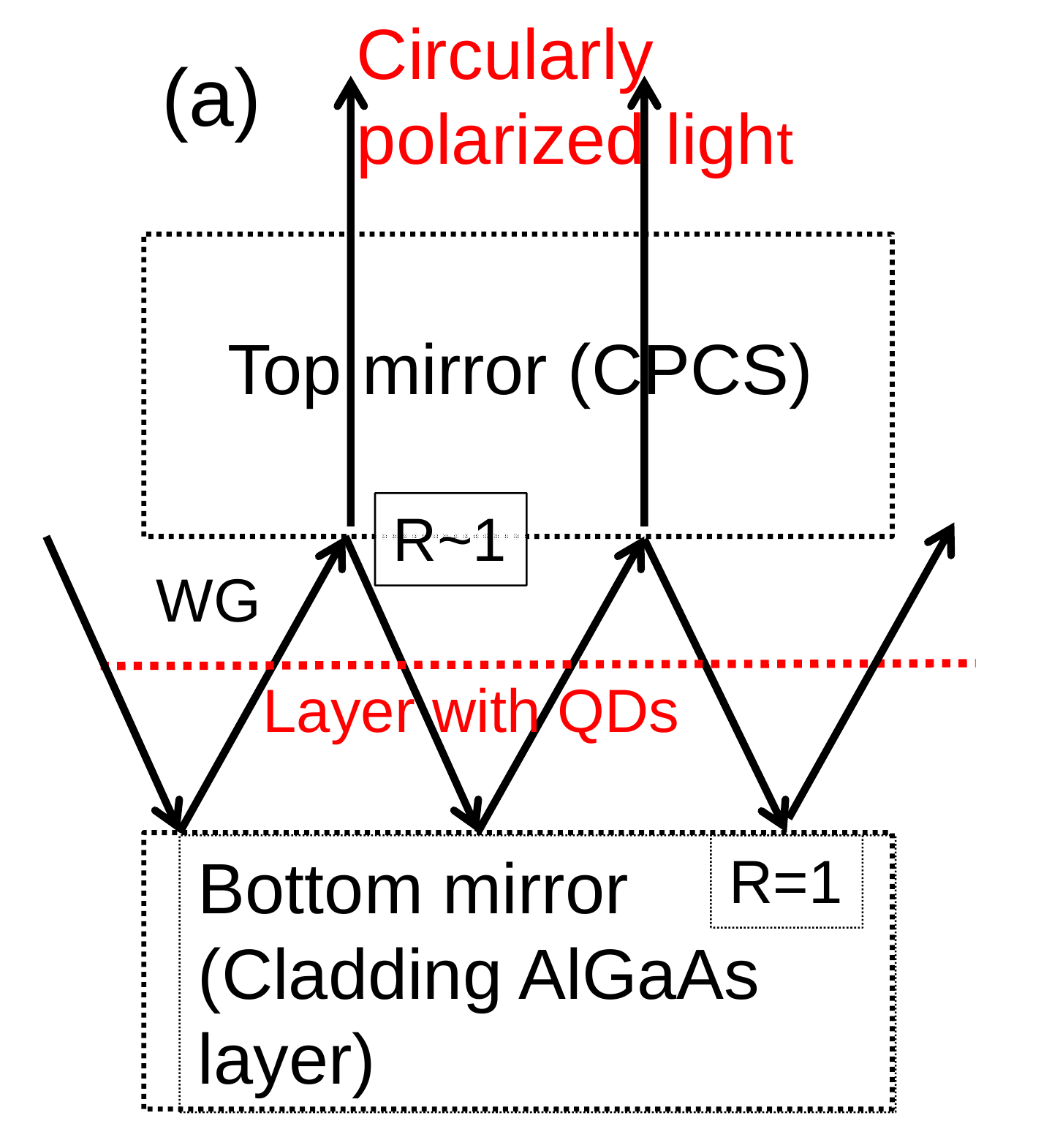}
\includegraphics[width=0.45\linewidth]{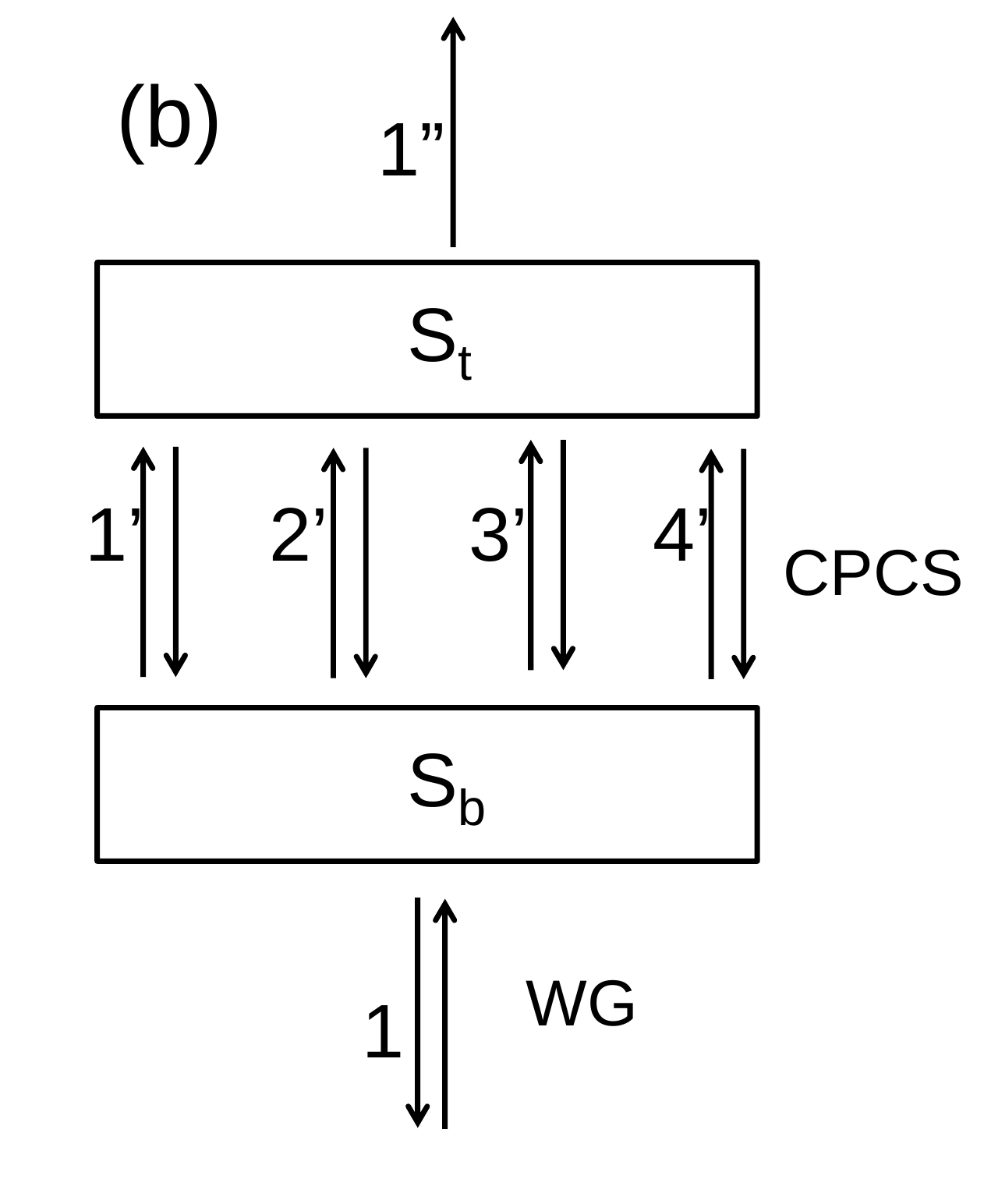}
\caption{ \label{fig_10} (a) Scheme of light scattering in the structure.
(b) Schematics of $\mathbf{k}=0$ light emission near the $\lambda=910.5$~nm resonance. Only the propagating modes are shown,
with wavenumbers (in units of $k_0$) $K_{z1''}=1$; $K_{z1',2',3',4'}=2.61, 1.85, 1.32, 1.22$;
$K_{z1}=1.11$. Actually, there are 12 more propagating modes in the waveguide layer (including the $k=0, K_{z}/k_0=\sqrt{\varepsilon}=3.52$ mode),  but in comparison with the
quasiguided mode 1 they do not scatter back from the bottom interface of the waveguide even if they are excited on the top interface and thus can be neglected.
}
\end{figure}

The optical scattering matrix can be approximately replaced by the unitary scattering matrix $S_u$.\cite{Gippius2005} The thicker the chiral photonic
crystal layer is, the better is this approximation. The schematics of the QD emission near the $\lambda_{0,\mathrm{th}}=910.5$~nm resonance
 with $n = +1$ or $-1$ (decoupled in case of C$_4$ symmetry) is shown in Fig.~\ref{fig_10}a.

\begin{figure}[t]
\includegraphics[width=0.99\linewidth]{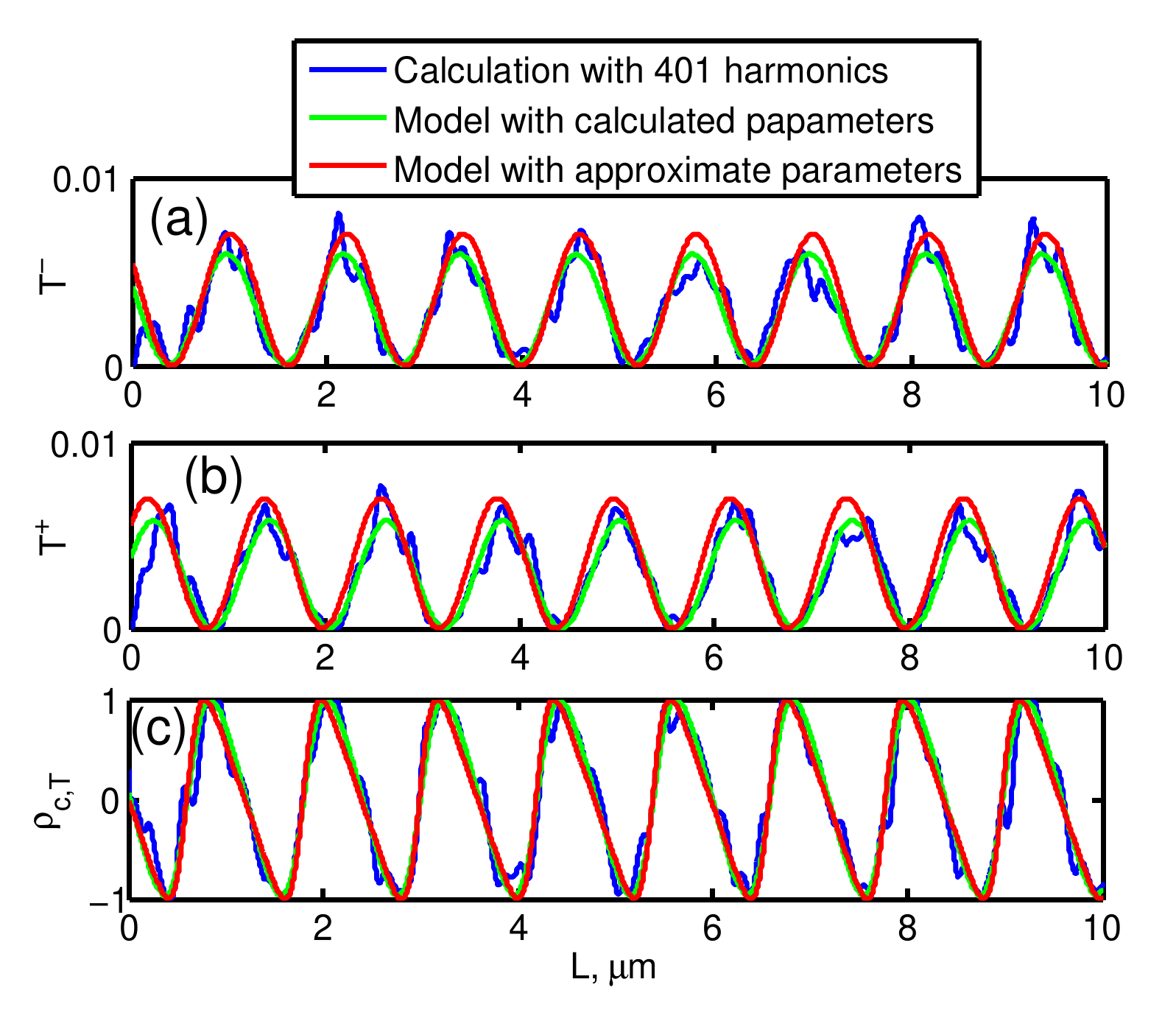}
\caption{ \label{fig_11} (Color online)  The calculated dependencies of the CPCS transmission $T^\pm$ in (a) right and (b) left  polarizations on CPCS thickness $L$. Panel (c) shows
the calculated circular selectivity of  transmission $\rho_{c,T}$. Blue lines are the results of calculations
with 401 spatial harmonics. Green and red lines correspond to the two-modes model Eq.~\ref{Tpmsimple2} with
approximated and calculated (using 401 harmonics) parameters (see in Tab.~\ref{tab}), respectively.
}
\end{figure}
Here, $S_t$ and $S_b$ are the unitary scattering matrixes of the air-CPCS and CPCS-waveguide interfaces, respectively. Hereinafter we omit
the subscript $u$ in the scattering matrix notation for simplicity.

In general, the unitary scattering matrix expresses the amplitudes of outgoing modes in terms of the amplitudes of incoming modes:
\begin{equation}\label{Smatrix}
    \left(
   \begin{array}{c}
     \mathrm{out}_1 \\
     \mathrm{out}_2 \\
     \mathrm{out}_3 \\
     \cdots
   \end{array}
   \right)
   =
   S
       \left(
      \begin{array}{c}
     \mathrm{in}_1 \\
     \mathrm{in}_2 \\
     \mathrm{in}_3 \\
     \cdots
   \end{array}
   \right)
   =
          \left(
      \begin{array}{cccc}
     r_1       & t_{12}  & t_{13}  & .\\
     t_{21}   &  r_2     & t_{23}  & .\\
      t_{31}  & t_{32}  & r_3      & .\\
    \cdots   & \cdots  & \cdots & .
   \end{array}
   \right)
          \left(
      \begin{array}{c}
     \mathrm{in}_1 \\
     \mathrm{in}_2 \\
     \mathrm{in}_3 \\
     \cdots
   \end{array}
   \right).
\end{equation}
The diagonal elements of the unitary scattering matrix are the reflection amplitudes $r_i$ of the $i^\mathrm{th}$ channel, and non-diagonal elements
are the transition amplitudes $t_{ij}$ from the $j^\mathrm{th}$ channel to the $i^\mathrm{th}$ channel.

Similar to the relations in Eq.~\ref{pm}, the time-reversal symmetry of Maxwell's equations for systems consisting from
 non-absorbing and non-gyrotropic materials relates also reflection and transition amplitudes in $\sigma^\pm$ polarizations:
\begin{equation}\label{rrtt}
 r_i^+ = r_i^- , t_{ij}^+ = t_{ji}^-.
\end{equation}

\subsubsection{\label{Subsubsec4.2.4} Two-mode resonances}

In the structure of interest, at the resonance wavelength, the QDs excite mode $\#1$, which leaks along $z$ axis primarily through
the channels $\#1'$ and $\#2'$ with transition amplitudes $t_{1\rightarrow1'}^\pm$ and $t_{1\rightarrow2'}^\pm$, respectively, see Fig.~\ref{fig_10}b.
For both polarizations $\sigma^+$ or $\sigma^-$, these modes propagate through the CPCS with wavenumbers $K_{z1'}$ and $K_{z2'}$,
respectively. At the top boundary of the CPCS these modes transform into the mode $\#1''$ with the transition amplitudes $t_{1'\rightarrow1''}^\pm$
and $t_{2'\rightarrow1''}^\pm$ , respectively. The total transition probability, including the phase shift due to propagation over the CPCS, is
\begin{equation}\label{Tpm}
  T^\pm =\left|
   t^\pm_{1\rightarrow 1'} \cdot
   t^\pm_{1'\rightarrow1''}\cdot
  \mathrm{e}^{iK_{z1'}L}
  +
     t^\pm_{1\rightarrow 2'}\cdot
   t^\pm_{2'\rightarrow1''}\cdot
  \mathrm{e}^{iK_{z2'}L}
\right|^2 ,
\end{equation}
where $L$ is the thickness of the CPCS. To simplify the formulas, let's introduce the new variables
\begin{equation}\label{Tpm0}
T^\pm_0 = 4 |
 t^\pm_{1\rightarrow 1'} \cdot
   t^\pm_{1'\rightarrow1''}|^2     ,
\end{equation}
\begin{equation}\label{etapm}
\eta^\pm = \left| \frac{
 t^\pm_{1\rightarrow 2'} \cdot
   t^\pm_{2'\rightarrow1''}
   }{
 t^\pm_{1\rightarrow 1'} \cdot
   t^\pm_{1'\rightarrow1''}
}
   \right|
\end{equation}
\begin{equation}\label{phipm}
\varphi^\pm = \arg \left( \frac{
 t^\pm_{1\rightarrow 2'} \cdot
   t^\pm_{2'\rightarrow1''}
   }{
 t^\pm_{1\rightarrow 1'} \cdot
   t^\pm_{1'\rightarrow1''}
}
   \right),
\end{equation}
$\Delta K =K_{z2'}-K_{z1'}$. Then
\begin{equation}\label{Tpmsimple1}
    T^\pm = \frac{T^\pm_0}{4}
    \left|
1
  +
     \eta^\pm
  \mathrm{e}^{i\Delta KL+\varphi^\pm}
\right|^2 .
\end{equation}
If $\eta^\pm = 1$, then
\begin{equation}\label{Tpmsimple2}
    T^\pm = T^\pm_0
    \cos ^2\frac{\Delta KL+\varphi^\pm}{2}
\end{equation}
is the oscillating function of the thickness $L$, which varies from 0 to $T_0^\pm$.
Moreover, if $|\varphi^+-\varphi^- = \pi$, the transmission in one polarization reaches its maximum
when the transmission in the opposite polarization reaches minimum and vice versa. The circular polarization  selectivity of transmission $\rho_{c,T}=(T^+-T^-)/(T^++T^-)$ in
these thicknesses appears to be around 100\%.

A simple model to estimate the introduced above parameters for the resonance at $E_0$  is given in the Appendix. The results of the simplified
model are compared with the optical scattering matrix calculation in Fig.~\ref{fig_11}. Although the parameters of the simplified model seem to differ substantially from the
ones calculated within the scattering matrix method with 401 harmonics (see in Tab.~\ref{tab}), the results from this crude model
for the circular polarization selectivity of transmission (panel c in Fig.~\ref{fig_11}) are in a good
quantitative agreement with the calculations within the scattering matrix. It happens because $\eta^\pm, |\varphi^+-\varphi^-|$ and especially the
$\rho_c$ are sensitive to the ratios of the partial transmissions trough modes $1'$ and $2'$,  which deviate less
from the calculated values.

The dependence $\rho_{c,T}(L)$ is also shown as blue line in panel b of Fig.~\ref{fig_09}. It can be seen that it is very close to the dependence  $\rho_{c}(L)$ (green line in the same
panel). This fact supports our assumption that it is the selectivity of transmission through the chiral photonic slab that controls the degree of circular polarization of QDs
emission in our structure.

\section{\label{Sec5} Conclusion}

In conclusion, we have fabricated and studied a structure with light emitting
InAs QDs inside a GaAs waveguide with a chiral photonic
crystal layer on top that demonstrates a strongly
circularly polarized photoemission of QDs at zero magnetic field with $|\rho_c|$ up to
81\% and up to 96\% if the unpolarized background due
to CPCS imperfection is removed.
It indicates that the predicted values of $|\rho_c|$ close to 100\% can be reached with
further progress in fabrication technology.

The measured angular dependence of the circularly polarized emission intensities
near the resonance is shown to agree with the calculations using the optical
scattering matrix. This allows to attribute the circularly polarized resonant emission
 to the excitation by QDs of the
$(\pm 2,\pm 2)$ Bragg resonance of quasiguided modes in the GaAs waveguide.
Based on the reciprocity and symmetry analysis of the structure, we show that the chiral photonic crystal slab in
the suggested waveguide structure works as a half-wave plate, exploring the
Fabry-Perot interference between the propagating modes
in CPCS, to reach nearly a 100\% circular polarization of
the transmission. The advantage of using the  photonic crystal slab with a large contrast
of dielectric permittivities  is its giant chirality, which allows
to fabricate a very thin half-wave plate, with a thickness of the order of the emitted light wavelength.
Additional advantage of the CPCS wave plate is its in-plane rotational isotropy, due to the C$_4$ symmetry.

\begin{acknowledgments}
This work has been funded by
the project SPANGL4Q, under FET-Open grant number
FP7-284743,
and
RFBR Projects No. 13-02-12144 and 14-02-00778.
We are grateful to K. Konishi,
L. Kuipers, M. Kuwata-Gonokami, R. Oulton, and H. Tamaru  for fruitful discussions.
We thank M. Emmerling for preparing the nanopillars, and acknowledge support by the state of Bavaria as well as the MWK Baden-
W\"{u}rttemberg.
\end{acknowledgments}

\appendix

\section{Analytical estimation of transition amplitudes}

Here we investigate the resonance shown in Fig.~\ref{fig_04} in the optimized structure and assume the free space wavelength
of light to be  fixed and equal to 910.5~nm.
The transition amplitudes $t_{i\rightarrow j}$ in Eqs.~(\ref{Tpm0})-(\ref{phipm}) can be roughly estimated via
 the normalized (see after Eq.~(\ref{pm})) in-plane components of the electric and magnetic fields of  modes $i,j$ as
\begin{equation}\label{ApprTrans}
  t_{i\rightarrow j} = \iint_{-p/2}^{p/2}\mathrm{d}x\mathrm{d}y\frac{
  \mathbf{E}_j^*\times\mathbf{H}_i + \mathbf{E}_i\times\mathbf{H}_j^* }{2}.
\end{equation}
Here, the symbol $\times$ means the pseudoscalar product
\begin{equation}
\mathbf{E}\times\mathbf{H}=E_xH_y-E_yH_x.
\end{equation}

The CPCS consists of dielectric rectangular pillars that are known to support the waveguide modes (in the vertical direction).
The electric (magnetic) field distribution of the waveguide mode
with the larger/smaller value of wavenumber $K_z$ can be roughly approximated by a homogeneous field parallel to the longest/shortest (shortest/longest) side inside the pillar
and zero field outside it.
Considering four rectangular pillars each rotated by $90^\circ$ with respect to its nearest neighbors requires the electric and magnetic fields
to satisfy Eq.~(\ref{Fxy}).
Note, due to the condition~(\ref{Fxy}), the integral in Eq.~\ref{ApprTrans} can be taken only over
 the first quarter of the elementary unit cell (it should be also multiplied by~4), since $\mathbf{E}^*(-y,x)\times \mathbf{H}(-y,x) = \mathbf{E}^*(x,y)\times \mathbf{H}(x,y)$.

The electric and magnetic field distributions of modes $\#1'$ and $\#2'$ in the first quarter of the elementary unit cell are
\begin{gather}\label{E1'}
     \mathbf{E}_{1'}^\pm(x,y)=
     \left\{
     \begin{array}{ll}
       \mp i C_1 \varepsilon^{-1/4}\mathbf{e}_y, & \mathrm{if}~x \in \left(\frac{p}{5},\frac{2p}{5} \right) \&~y \in \left(0,\frac{2p}{5} \right) \\
       0 , &  \mathrm{otherwise}      ,
     \end{array}
     \right.\\
     \mathbf{H}_{1'}^\pm(x,y)=
     \left\{
     \begin{array}{ll}
       \pm i C_1 \varepsilon^{1/4}\mathbf{e}_x, & \mathrm{if}~x \in \left(\frac{p}{5},\frac{2p}{5} \right) \&~y \in \left(0,\frac{2p}{5} \right) \\
       0 , &  \mathrm{otherwise}      ,
     \end{array}
     \right.\\
     \mathbf{E}_{2'}^\pm(x,y)=
     \left\{
     \begin{array}{ll}
       C_1 \varepsilon^{-1/4}\mathbf{e}_x, & \mathrm{if}~x \in \left(\frac{p}{5},\frac{2p}{5} \right) \&~y \in \left(0,\frac{2p}{5} \right) \\
       0 , & \mathrm{otherwise}        ,
     \end{array}
     \right.\\
     \mathbf{H}_{2'}^\pm(x,y)=
     \left\{
     \begin{array}{ll}
       C_1 \varepsilon^{1/4}\mathbf{e}_y, & \mathrm{if}~x \in \left(\frac{p}{5},\frac{2p}{5} \right) \&~y \in \left(0,\frac{2p}{5} \right) \\
       0 , & \mathrm{otherwise}        ,
     \end{array}
     \right.
\end{gather}
where the constant $C_1$ is introduced for brevity,
$$
     C_1 = \frac{5}{2\sqrt{2}p}.
$$

The electric and magnetic field distributions of the modes $\#1$ (which is the $(\pm 2, \pm 2)$ Bragg resonance of the GaAs layer guided mode, in
the horizontal direction) and $\#1''$ are
 \begin{gather}
    \mathbf{E}_1^\pm(x,y) = \frac{1}{p}\sqrt{\frac{k_0}{K_{z1}}} \nonumber \\
    \left[
    \frac{-\mathbf{e}_x+\mathbf{e}_y}{\sqrt{2}}
    \cos \left(
    \frac{4 \pi}{p}(x+y)
      \right)
    \pm i
 \frac{\mathbf{e}_x+\mathbf{e}_y}{\sqrt{2}}
    \cos \left(
    \frac{4 \pi}{p}(x-y)
\right)
    \right],\\
    \mathbf{H}_1^\pm(x,y) = \frac{1}{p}\sqrt{\frac{K_{z1}}{k_0}} \nonumber \\
    \left[
    -\frac{\mathbf{e}_x+\mathbf{e}_y}{\sqrt{2}}
    \cos \left(
    \frac{4 \pi}{p}(x+y)
      \right)
    \pm i
 \frac{-\mathbf{e}_x+\mathbf{e}_y}{\sqrt{2}}
    \cos \left(
    \frac{4 \pi}{p}(x-y)
\right)
    \right],\\
\mathbf{E}_{1''}^\pm(x,y) =  \frac{1}{p}  \frac{\mathbf{e}_x \mp i \mathbf{e}_y}{\sqrt{2}},\\
\label{H1''}
\mathbf{H}_{1''}^\pm(x,y) =  \frac{1}{p}  \frac{\mathbf{e}_y \pm i \mathbf{e}_x}{\sqrt{2}},
\end{gather}
where the wavenumber is
\begin{equation}\label{Kz1}
K_{z1}=\sqrt{\varepsilon k_0^2-\frac{32\pi^2}{p^2} },
\end{equation}
 $\varepsilon = 12.42$ is GaAs dielectric permittivity.

Substituting (\ref{E1'})-(\ref{H1''}) in (\ref{ApprTrans}), one can find the transition amplitudes
\begin{gather}\label{t1}
  t_{1\rightarrow1'}^\pm = C_2
  \left( -C_3 \pm
i C_4
  \right) ,\\
  t_{1\rightarrow2'}^\pm = C_2 \left(
-C_4
\pm i C_3
  \right) ,\\
\label{tt}
      t_{1'\rightarrow1''}^\pm =  t_{2'\rightarrow1''}^\pm =  C_5,
\end{gather}
where the constants $C_2$, $C_3$, $C_4$, and $C_5$ are
\begin{gather}
  C_2 = \frac{5}{16\pi^2}
  \left(\varepsilon^{1/4}\sqrt{\frac{k_0}{K_{z1}}} +
  \varepsilon^{-1/4}\sqrt{\frac{K_{z1}}{k_0}}\right) , \\
  C_3 = \sin^2\frac{\pi}{5} ,\\
  C_4 = \cos\frac{2\pi}{5} + \cos\frac{\pi}{5} ,\\
  C_5 = \frac{1}{5}\left(\varepsilon^{1/4}+\varepsilon^{-1/4}\right) .
\end{gather}

\begingroup
\squeezetable
\begin{table}[b]
\caption{\label{tab}
Comparison of the simple model in Appendix with the numerical simulation with 401 harmonics}
\begin{ruledtabular}
\begin{tabular}{c c c c}									
Variable	 						& Approximate 			&	Simulation	&    Relative difference
\footnote{Here, we calculate the relative difference $\delta$ of two variables $a$ and $b$ as $\delta=\frac{|a-b|}{|a|+|b|}$.}	  \\
\hline	
$t_{1\rightarrow1'}^+$		&  -0.0256 + 0.0830i  	& 0.0093 + 0.0673i      &  25\%  \\
$t_{1\rightarrow1'}^-$      &  -0.0256 - 0.0830i  	& -0.0487 - 0.0564i     &  22\%  \\
$t_{1\rightarrow2'}^+$		&  -0.0830 + 0.0256i 	& -0.0452 + 0.0220i     &  28\%  \\
$t_{1\rightarrow2'}^-$      &  -0.0830 - 0.0256i  	& -0.0417 + 0.0199i     &  46\%  \\
$t_{1'\rightarrow1''}^+$	&  0.4820  				& 0.5774 - 0.0613i      &  11\%  \\
$t_{1'\rightarrow1''}^-$    &  0.4820 				& 0.5734 - 0.0994i      &  13\%  \\
$t_{2'\rightarrow1''}^+$	&  0.4820 				& 0.7290 - 0.0561i      &  21\%  \\
$t_{2'\rightarrow1''}^-$    &  0.4820 				& 0.7277 - 0.0122i      &  20\%  \\
$T_0^+$		                &  0.007 				& 0.0062      			&  6\%  \\
$T_0^-$                     &  0.007 				& 0.0075       			&  4\%  \\
$\eta^+$	                &  1					& 0.932        			&  4\%  \\
$\eta^-$                    &  1					& 0.775        			&  13\%  \\
$|\varphi^+-\varphi^-|$     &  111$^\circ$			& $139^\circ$	        &  11\%  \\
\end{tabular}
\end{ruledtabular}
\end{table}
\endgroup

Substituting formulas (\ref{t1})-(\ref{tt}) into the definitions of $T_0^\pm$, $\eta^\pm$, and $\varphi^\pm$, i.e., Eqs.(\ref{Tpm0}), (\ref{etapm}),
and (\ref{phipm}), respectively, one finds
\begin{gather}
T_0^\pm = C_2^2C_5^2(C_3^2+C_4^2)    \approx0.007,\\
 \eta^\pm=1, \\
|\varphi^+-\varphi^- |=\pi-4 \mathrm{atan}\frac{C_3}{C_4} \approx 111^\mathrm{o} .
\end{gather}
    Thus, as discussed above in Eq.~\ref{Tpmsimple2},  transmission in each polarization depends periodically on the CPCS thickness. As the function of thickness, it varies from 0 to 0.007 and is
    shifted by 111$^\circ$ with respect to the transmission in the opposite polarization. The comparison of this simple model  with the numerical
    simulation with 401 harmonics is given in the Tab.~\ref{tab}. The resulting dependencies of the CPCS transmission in $\sigma^+$ and $\sigma^-$ polarizations
    and of the circular polarization degree of transmission in comparison with the same formulas (but using exact parameters from Tab.~\ref{tab}) and
the results of simulation with 401 harmonics are given in Fig.~\ref{fig_11}. It can be seen that this very approximate and crude model is in a qualitative agreement
with the numerical simulations.
The dependencies in Fig.~\ref{fig_11}, calculated with 401 harmonics, demonstrate also sharp peaks, which are due to the influence of other electromagnetic modes in the CPCS,
not taken into account in this simple two-mode model.
More detail on the comparison of the model with numerical simulations are given in the Supplementary material.


%

\newpage
\begin{figure*}[htp] \centering{
\includegraphics[scale=0.85]{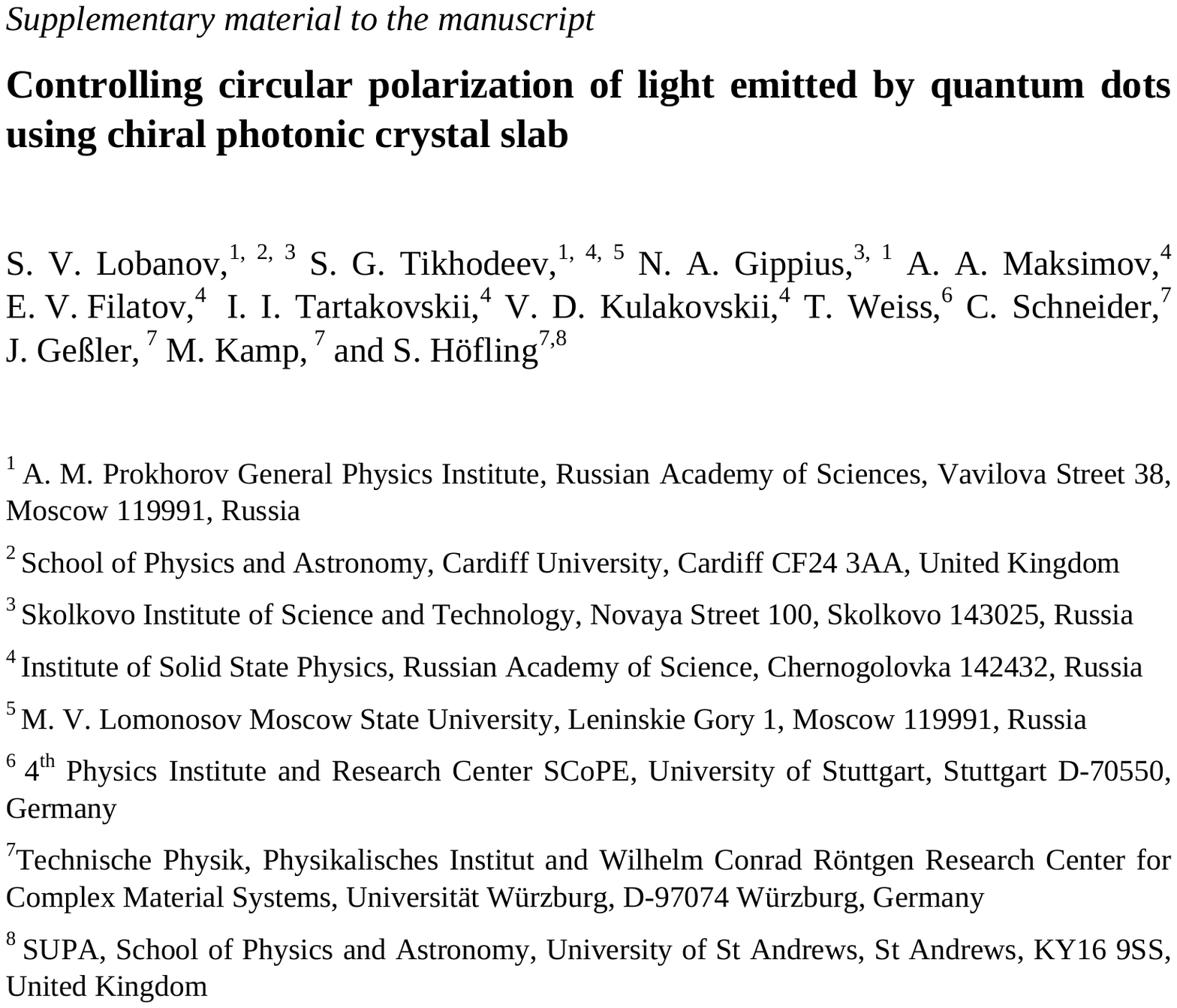}}
\end{figure*}

\begin{figure*}[htp] \centering{
\includegraphics[scale=0.85]{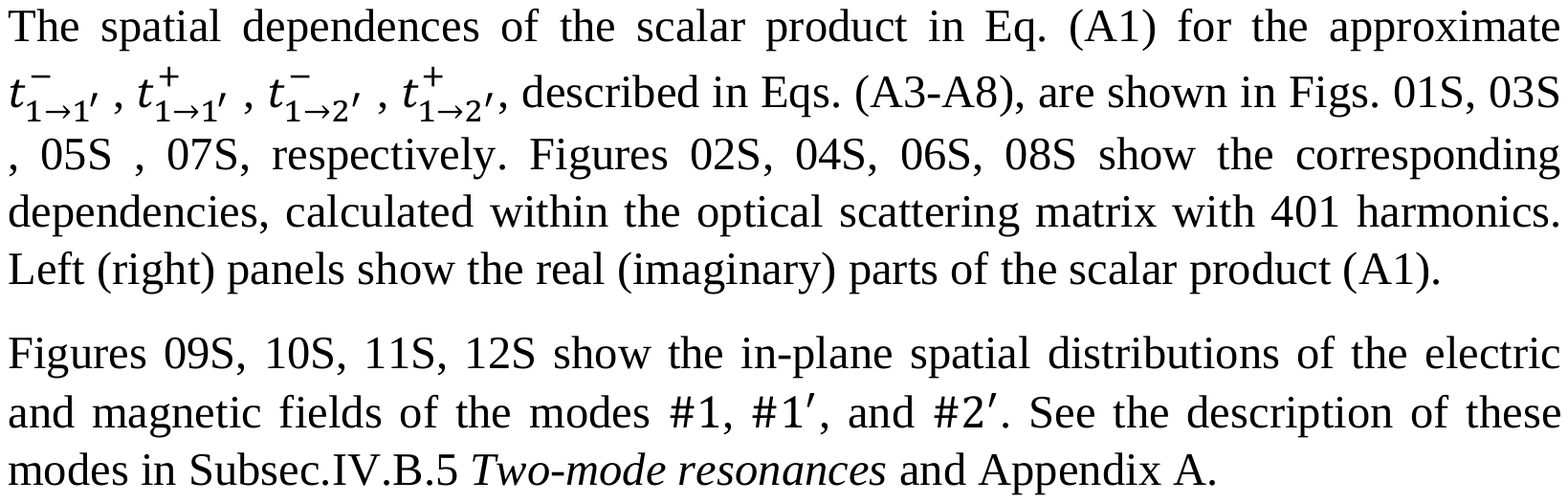}}
\end{figure*}

\begin{figure*}[htp] \centering{
\includegraphics[scale=0.85]{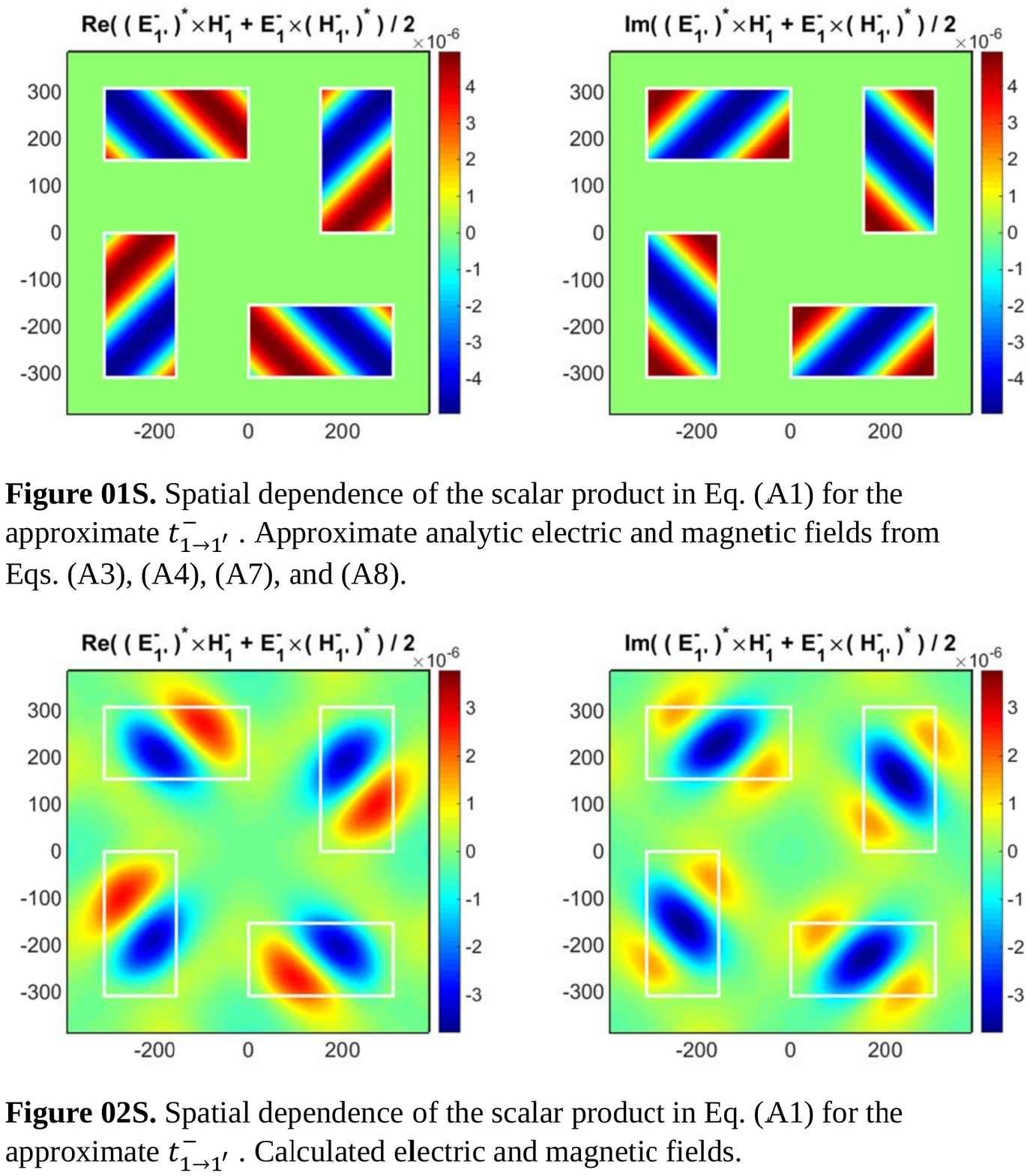}}
\end{figure*}

\begin{figure*}[htp] \centering{
\includegraphics[scale=0.85]{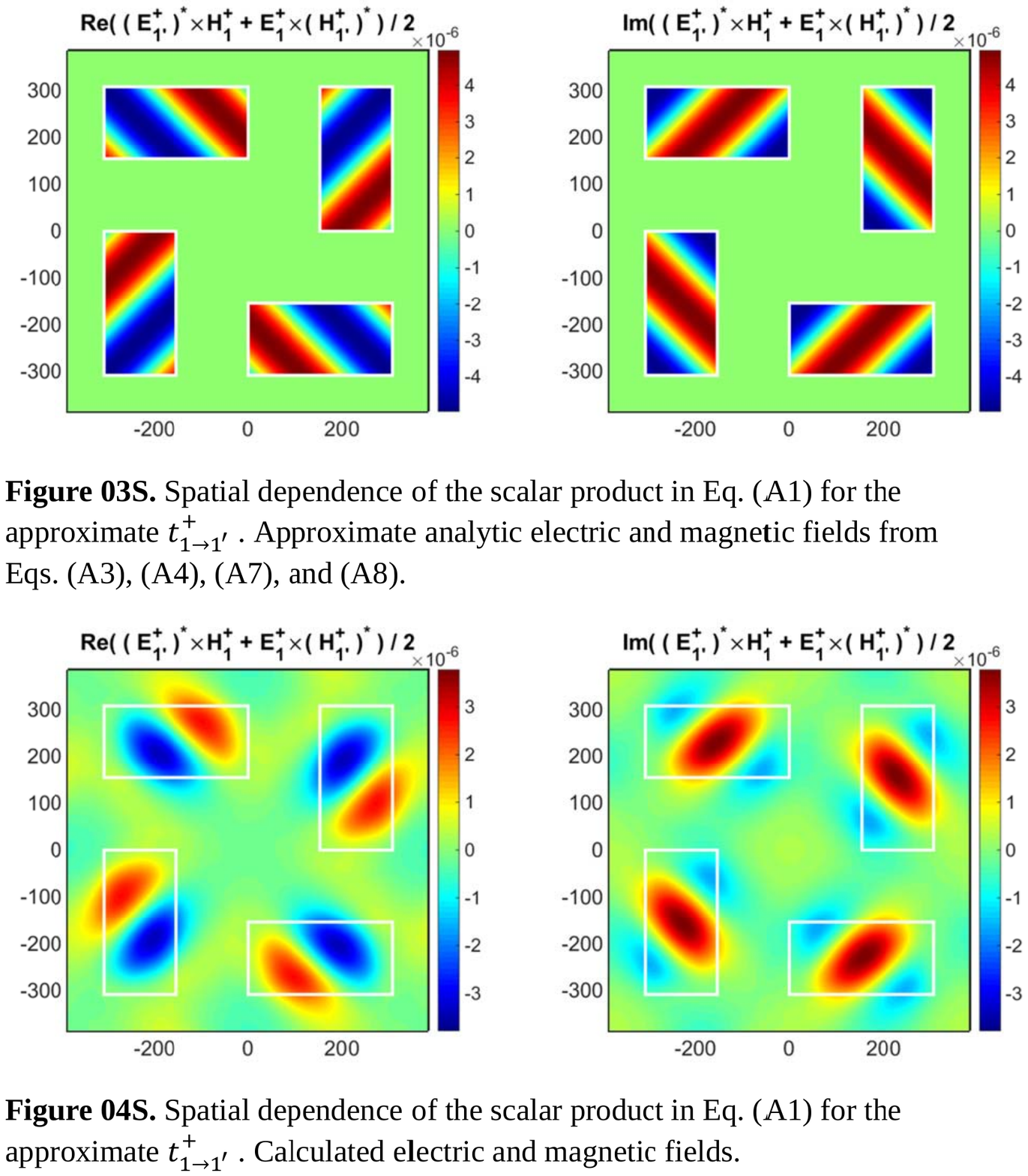}}
\end{figure*}

\begin{figure*}[htp] \centering{
\includegraphics[scale=0.85]{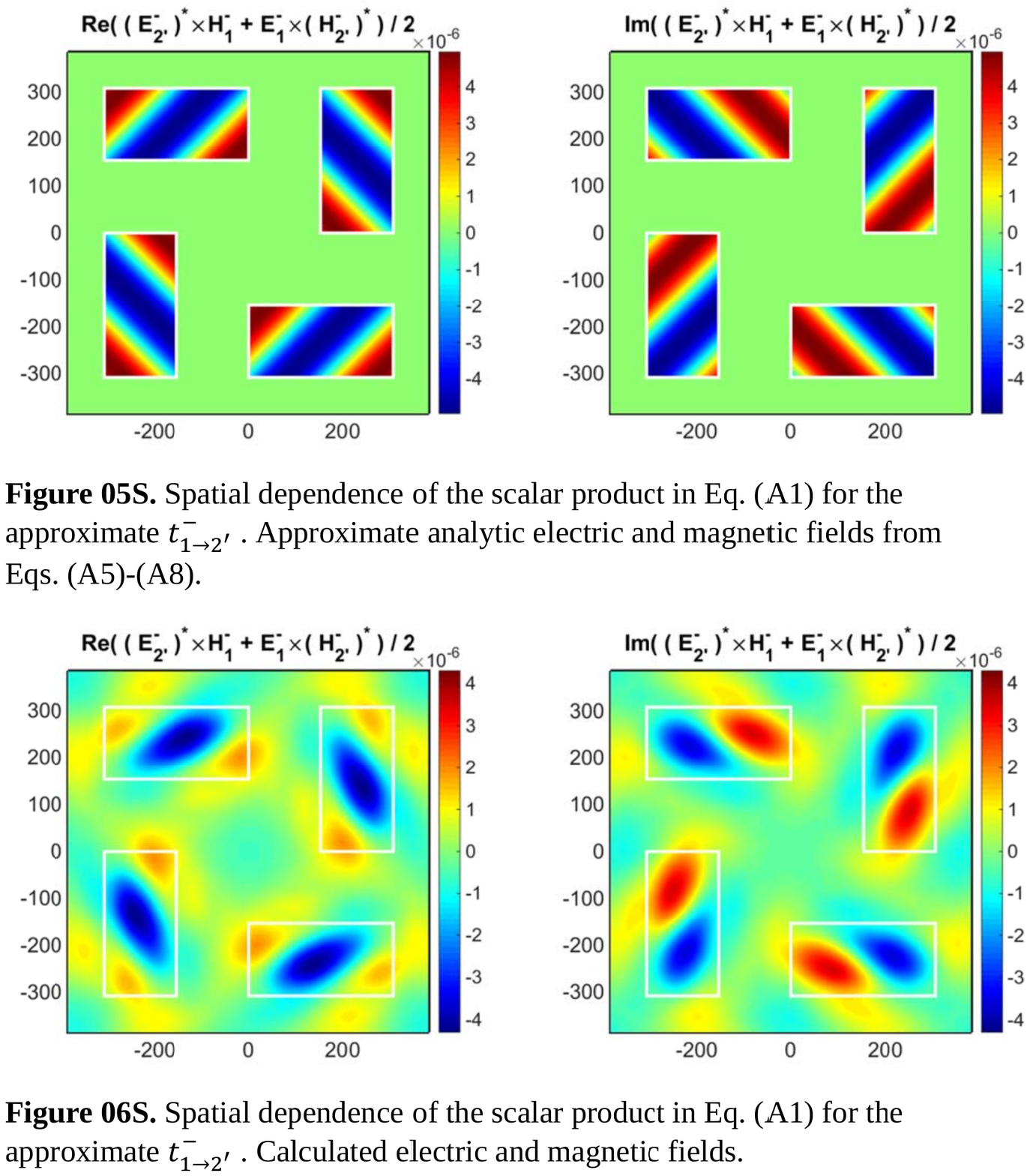}}
\end{figure*}

\begin{figure*}[htp] \centering{
\includegraphics[scale=0.85]{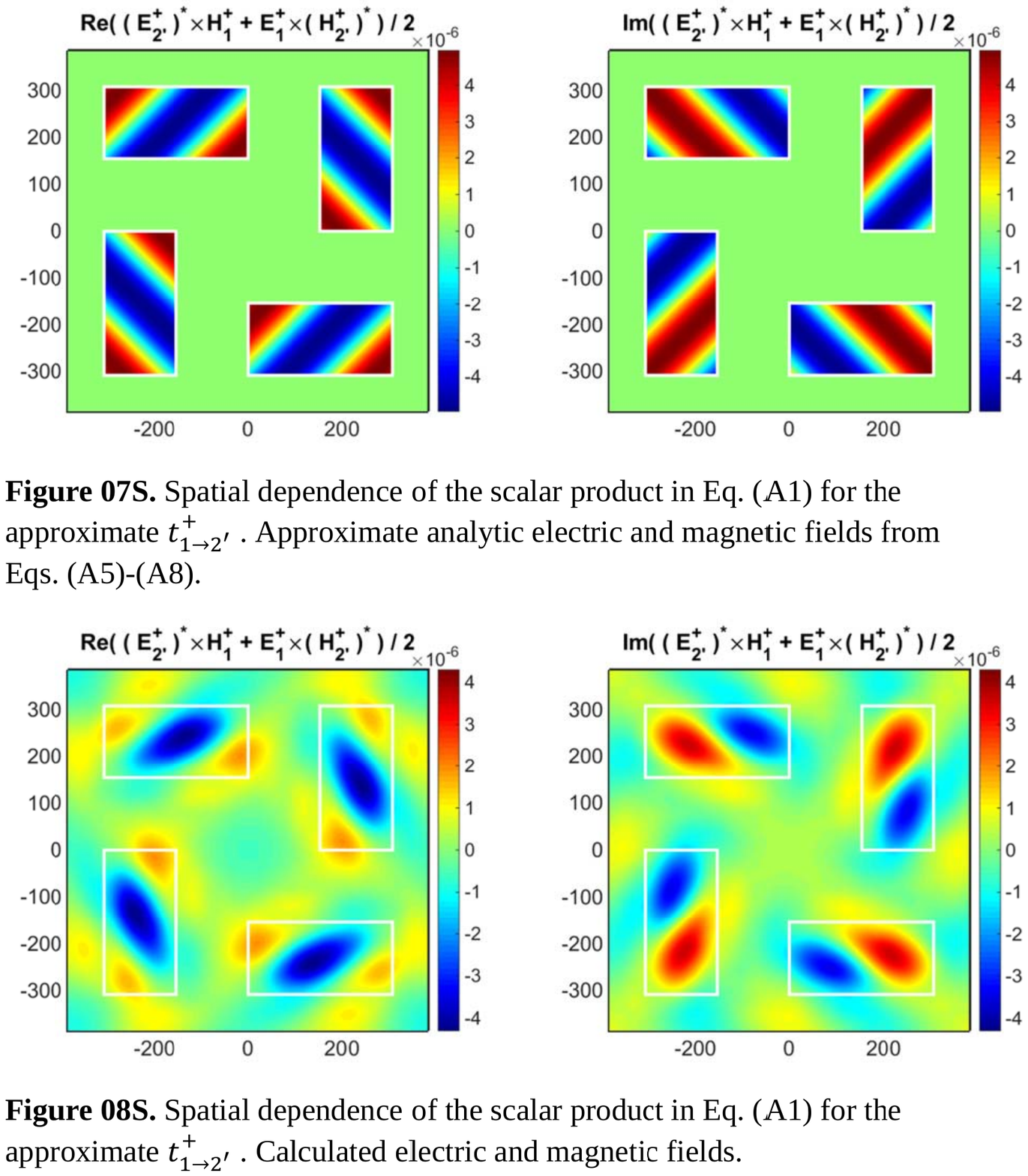}}
\end{figure*}

\begin{figure*}[htp] \centering{
\includegraphics[scale=0.85]{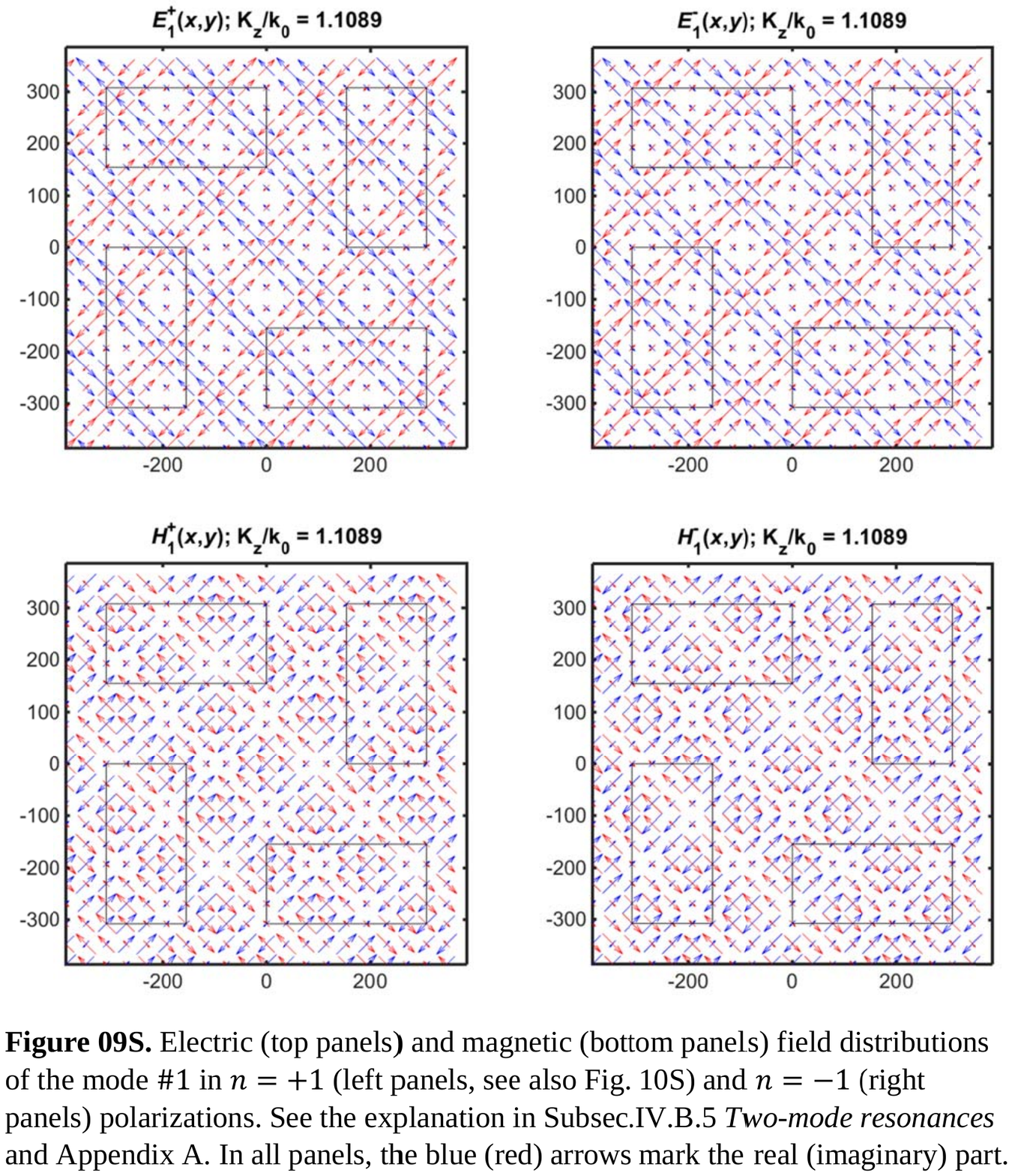}}
\end{figure*}

\begin{figure*}[htp] \centering{
\includegraphics[scale=0.85]{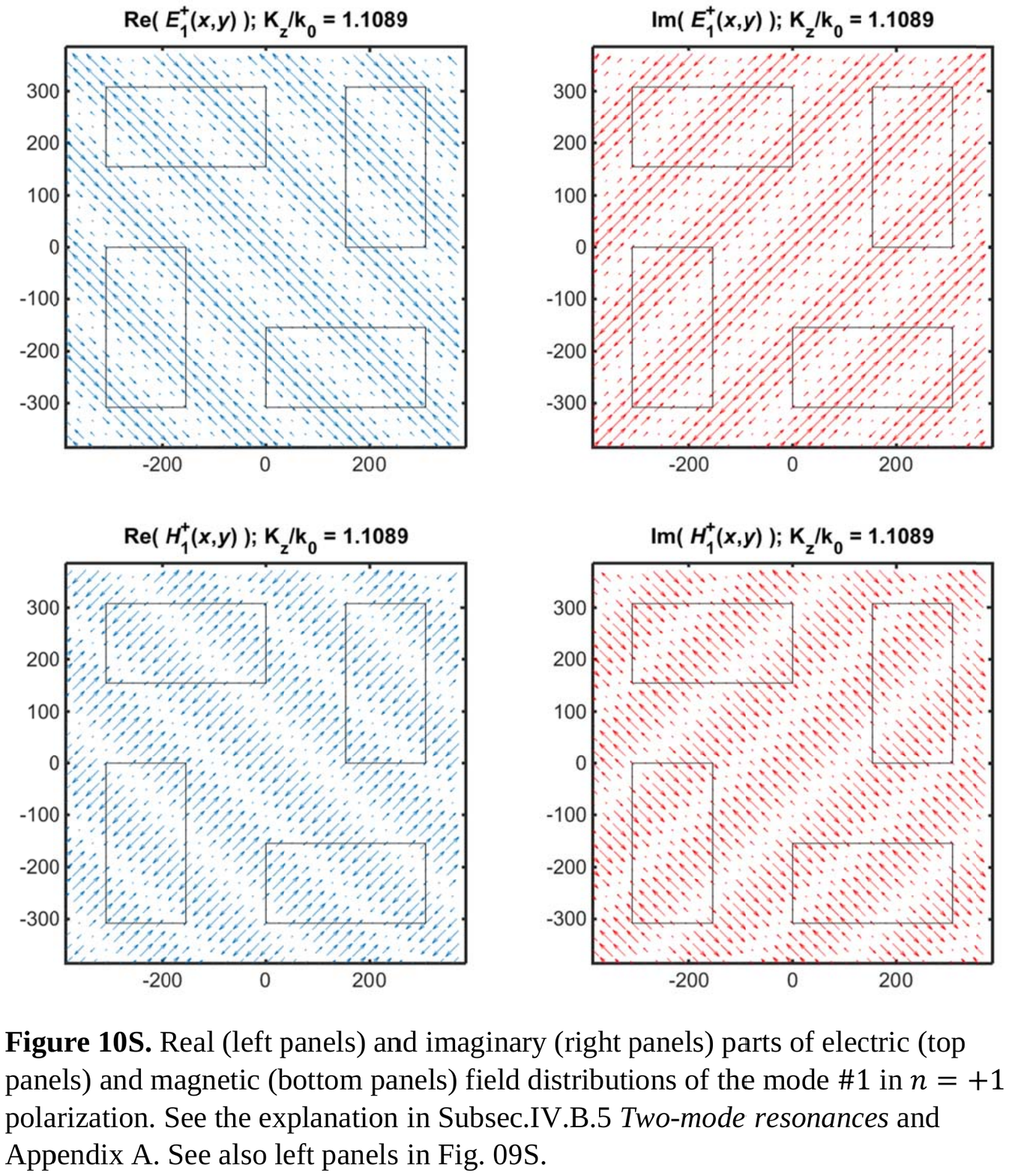}}
\end{figure*}

\begin{figure*}[htp] \centering{
\includegraphics[scale=0.85]{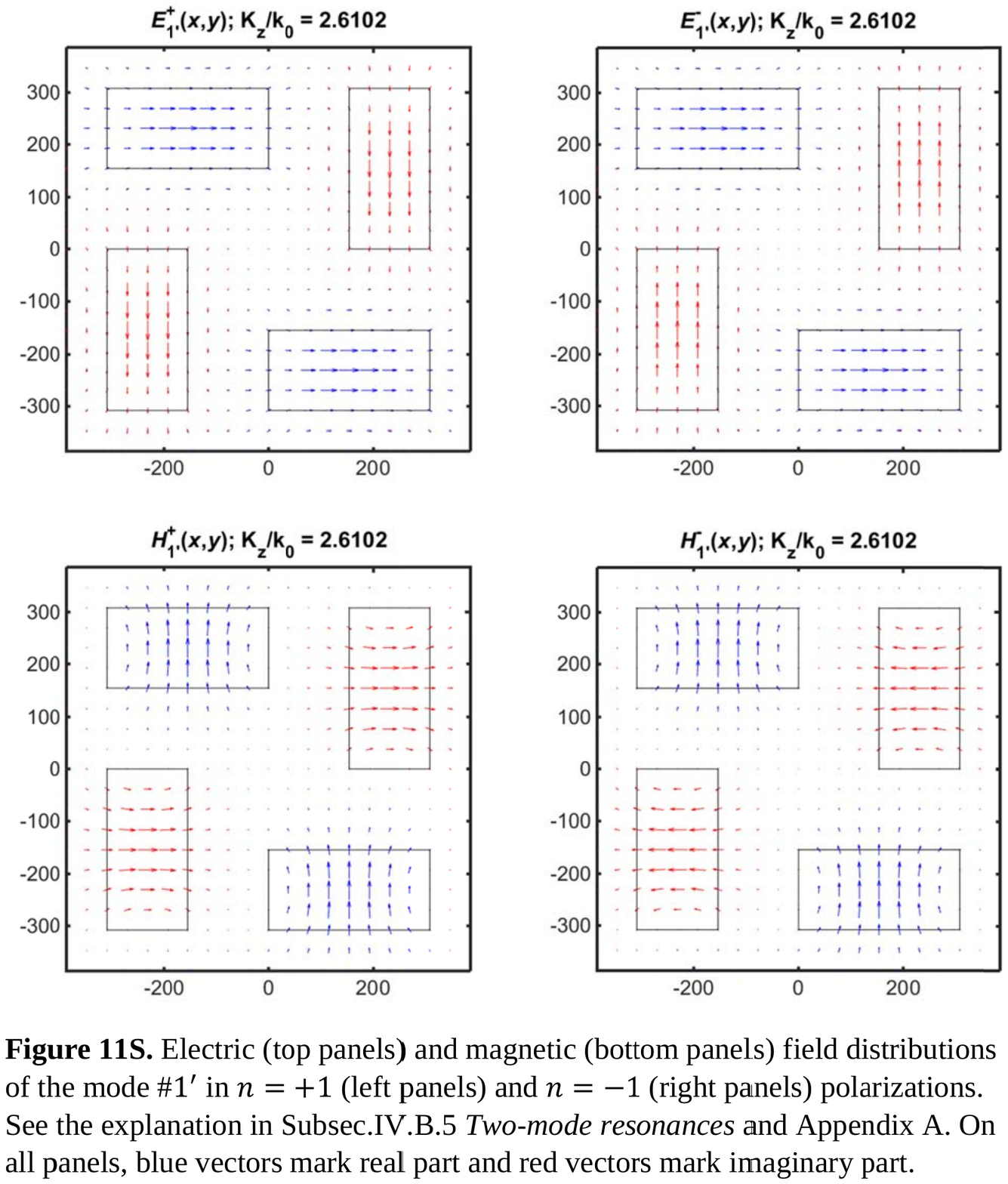}}
\end{figure*}

\begin{figure*}[htp] \centering{
\includegraphics[scale=0.85]{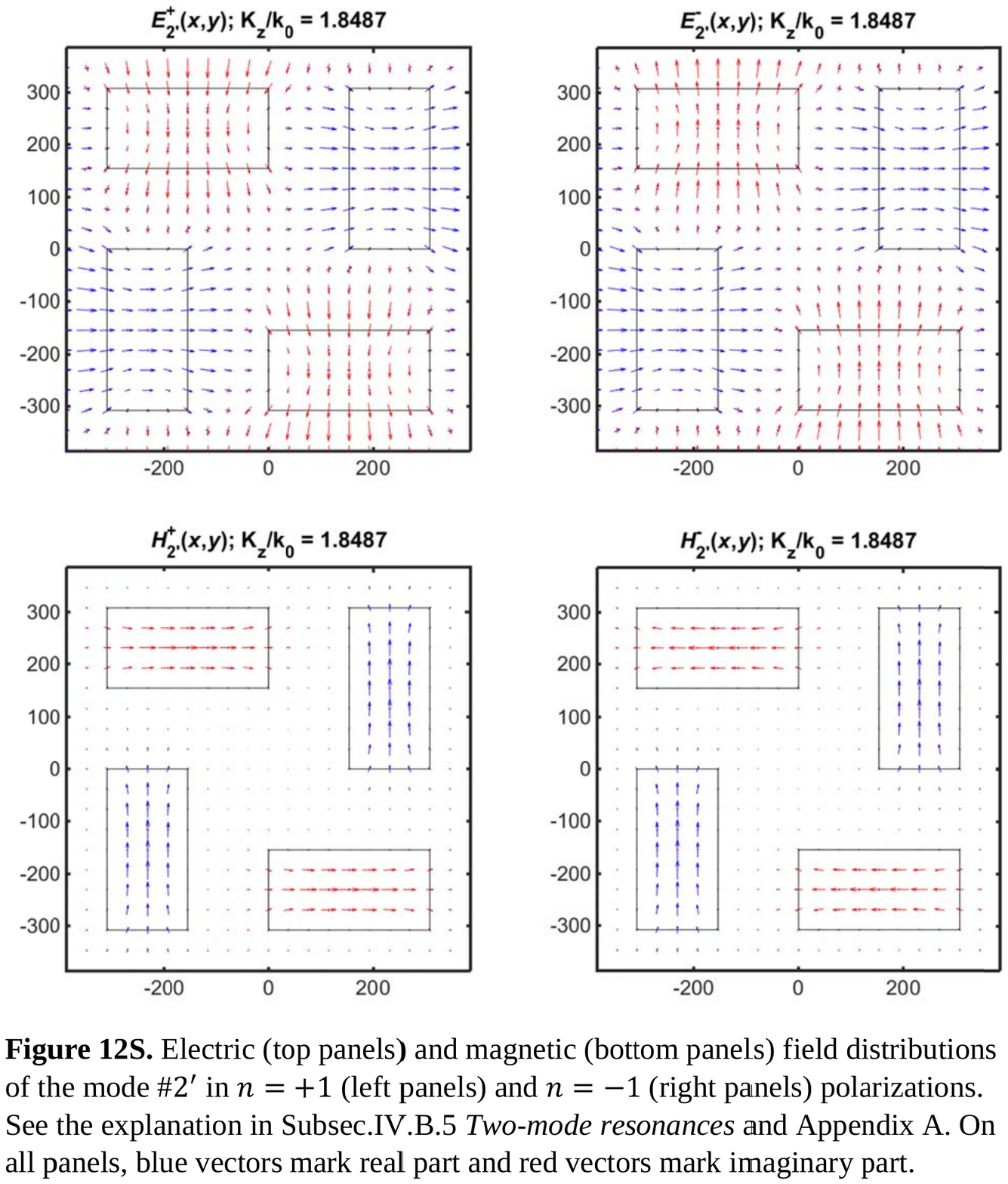}}
\end{figure*}

\end{document}